\documentclass[pra,reprint,showpacs,superscriptaddress]{revtex4-2}
\usepackage{graphicx}
\usepackage{bm}
\usepackage{amssymb}
\usepackage{amsmath}
\usepackage{amsthm}
\usepackage{times}
\usepackage[colorlinks=true,linkcolor=blue,urlcolor=blue,citecolor=blue,pdfauthor={ },pdftitle={ },pdfsubject={ },pdfkeywords={ }]{hyperref}
\graphicspath{{Figures/}}

\begin{document}
\title{Improved quantum computing with higher-order Trotter decomposition}
\author{Xiaodong Yang}
\affiliation{Shenzhen Institute for Quantum Science and Engineering, Southern University of Science and Technology, Shenzhen, 518055, China}
\affiliation{International Quantum Academy, Shenzhen, 518055, China}
\affiliation{Guangdong Provincial Key Laboratory of Quantum Science and Engineering, Southern University of Science and Technology, Shenzhen, 518055, China}

\author{Xinfang Nie}
\affiliation{Department of Physics, Southern University of Science and Technology, Shenzhen, P. R. China}
\affiliation{International Quantum Academy, Shenzhen, 518055, China}
\affiliation{Guangdong Provincial Key Laboratory of Quantum Science and Engineering, Southern University of Science and Technology, Shenzhen, 518055, China}

\author{Yunlan Ji}
\affiliation{School of Electronic Science and Applied Physics, Hefei University of Technology, Hefei, Anhui 230009, China}

\author{Tao Xin}
\affiliation{Shenzhen Institute for Quantum Science and Engineering, Southern University of Science and Technology, Shenzhen, 518055, China}
\affiliation{International Quantum Academy, Shenzhen, 518055, China}
\affiliation{Guangdong Provincial Key Laboratory of Quantum Science and Engineering, Southern University of Science and Technology, Shenzhen, 518055, China}

\author{Dawei Lu}
\email{ludw@sustech.edu.cn}
\affiliation{Department of Physics, Southern University of Science and Technology, Shenzhen, P. R. China}
\affiliation{Shenzhen Institute for Quantum Science and Engineering, Southern University of Science and Technology, Shenzhen, 518055, China}
\affiliation{International Quantum Academy, Shenzhen, 518055, China}
\affiliation{Guangdong Provincial Key Laboratory of Quantum Science and Engineering, Southern University of Science and Technology, Shenzhen, 518055, China}

\author{Jun Li}
\email{lij3@sustech.edu.cn}
\affiliation{Shenzhen Institute for Quantum Science and Engineering, Southern University of Science and Technology, Shenzhen, 518055, China}
\affiliation{International Quantum Academy, Shenzhen, 518055, China}
\affiliation{Guangdong Provincial Key Laboratory of Quantum Science and Engineering, Southern University of Science and Technology, Shenzhen, 518055, China}

\begin{abstract}
In designing quantum control, it is generally required to simulate the controlled system evolution with a classical computer. However,  computing the time evolution operator can be quite   resource-consuming    since   the total Hamiltonian is often hard to diagonalize.
In this paper, we mitigate this issue by substituting the time evolution segments with their Trotter decompositions, which reduces the propagator into a combination of single-qubit operations and fixed-time system evolutions. 
The resulting procedure can provide substantial speed gain with acceptable costs in the propagator error.  
As a demonstration, we apply the proposed strategy to improve the efficiency of the gradient ascent pulse engineering algorithm for searching optimal control fields.
Furthermore, we show that the higher-order Trotter decompositions can provide efficient Ans\"atze for the variational quantum algorithm, leading to improved performance in solving the ground-state problem. 
The strategy presented here is also applicable for many other quantum optimization and simulation tasks.
\end{abstract}
\date{\today}
\maketitle

\renewcommand{\thesubsection}{\arabic{subsection}}

\section{Introduction}
Quantum control offers access to explore various quantum phenomena and processes {\cite{WWS93,BCC10,koch2022quantum}}. {Normally, this is achieved by engineering a quantum system of interest to some control target with specifically designed time-dependent control fields {\cite{PRXQuantum.2.030203}.} Finding controls that allow optimal performance by analytical means is in general quite difficult \cite{DAD07}. For systems  with more than several qubits,  one often has to resort to  numerical approaches. In numerical quantum control, it is a routine to simulate the controlled system evolution on a classical computer and employ an iterative optimization algorithm  to search an optimal control \cite{KNR05,ZG16,YL19,RBS19}. During this procedure, the classical computer needs to compute a large number of matrix exponentials. Traditionally, there exist a variety of algorithms to compute matrix exponentials, while the most widely used variants are those that combine the Pad{\'e} approximants {\cite{MCV03}}. However, whichever algorithm is employed, this is generally a resource-consuming task, especially when the engineered Hamiltonian cannot be diagonalized. Current small-scale control optimizations can still be accomplished within an acceptable computer run time, yet it will  quickly become infeasible for problems of growing sizes in the coming noisy intermediate-scale quantum era \cite{PJN19}. Thus developing practical strategies to simulate large quantum system evolutions is of importance for quantum engineering on near-term quantum devices.

Previously, there have been proposed a number of   strategies to lower the difficulty of simulating controlled quantum dynamics. 
For example, compressing {the dimension of the evolution operator} with exquisite approximation techniques, like those used in the tensor-network-based framework \cite{WSR92,VGE03,EGG15,DPC} or in the subsystem-based  optimal control method \cite{RCN08,LJ19}, can substantially reduce the computational cost. However, these methods can only function in specific cases and need abundant pre-processing efforts. Additionally, the truncated Taylor series can be efficiently applied to approximate the evolution operator and decrease the computational complexity \cite{BDW15,KMS19}, yet it functions only when the controlled system Hamiltonian is sparse or a linear combination of some unitary terms. 
Alternatively, one can apply parallelization techniques to accelerate the computing process \cite{GTS06,ELJ12,RMK16}, though this requires special computer architectures and programming patterns. 
Recently, researchers attempt to partially ease the computational complexity with quantum resources, leading to hybrid quantum-classical control methods \cite{LYP17,LLL17}, which {need} extra experimental learning. Therefore, from the perspective of practicality, the above-mentioned strategies are not friendly for everyday use in the  laboratory.

Here, we put forward an effective and simple enough   strategy to improve the efficacy of simulating the system evolution  using the method of Trotter approximation \cite{JWS92,KHH94,HNS05}. The key ingredient of our strategy is to replace the matrix {exponential} of each time evolution segment with the Trotter decompositions.  As such, one only needs to compute the system evolution several times during the whole optimization process, which will shorten the computer run time largely. The parameters one has to optimize are then simply single-qubit rotations, which further reduces the computing time. However, one also faces a problem that the precision of the Trotterization decreases as the number of control parameters grows, so we propose to use the higher-order Trotter decompositions to get around the obstacle. A similar method has been explored in Ref. \cite{BJ18}. The numerical tests with the gradient ascent pulse engineering (GRAPE) algorithm for searching optimal control fields and the variational quantum algorithm (VQA) for solving the ground-state problem reveal the effectiveness of our strategy on reducing the computational resources.
  The outline of this paper is given as follows. We first introduce the proposed strategy in Sec. \ref{framework}, and the corresponding applications are presented in Sec. \ref{application}. Finally, we provide some brief discussions in Sec. \ref{discussion}.

\begin{figure}
\centering
\includegraphics[width=0.42\textwidth]{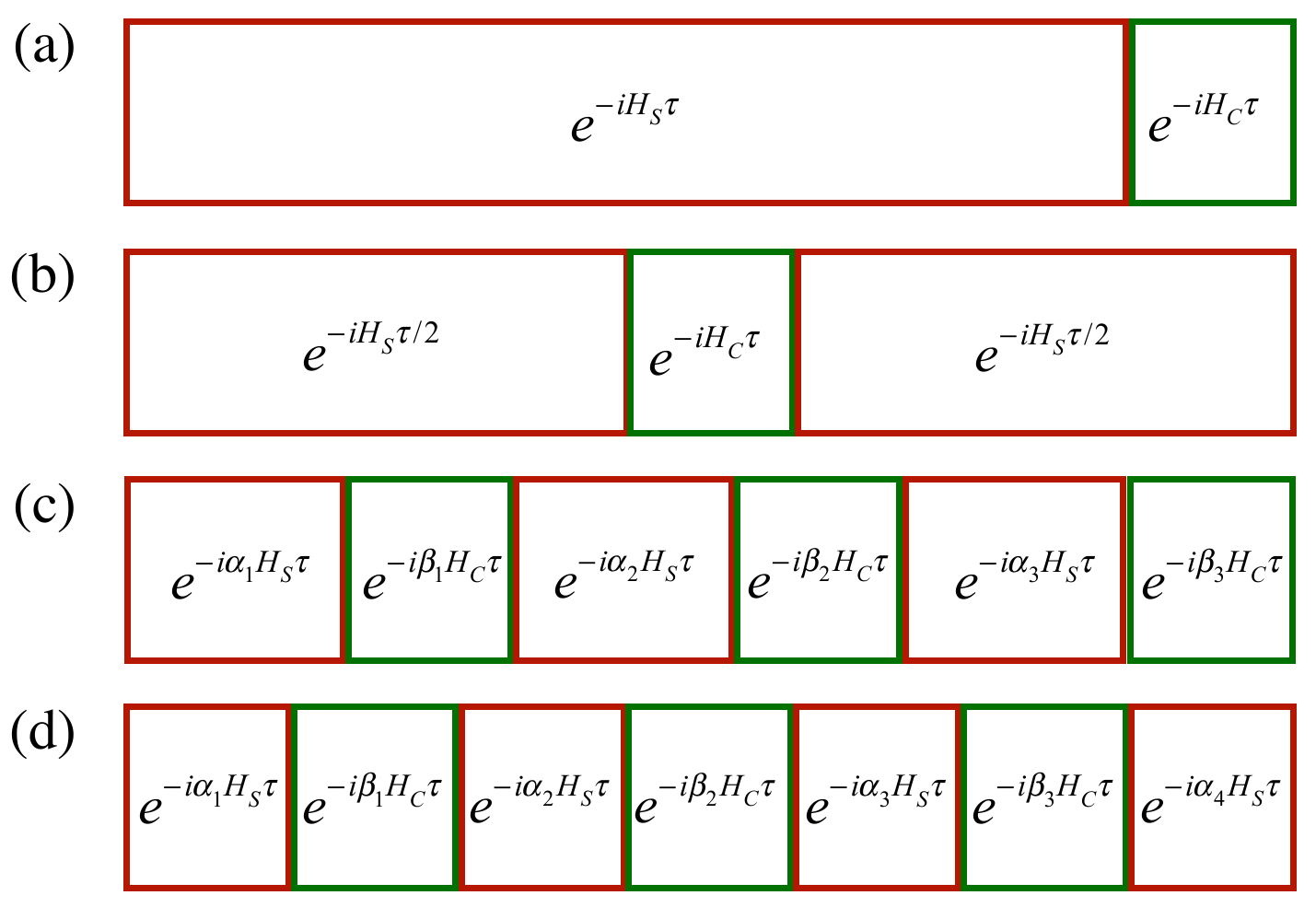}
\caption{Illustration of the sliced time evolution operator using different orders of the Trotter decomposition. (a) First order. (b) Second order. (c) Third order. The parameters are $\alpha_1=1-\gamma,\beta_1={(4/3-\gamma \pm \Gamma)}/[{2 \gamma}(\gamma \pm \Gamma)], \alpha_2=(\gamma \pm \Gamma)/2, \beta_2=(3-4 \gamma)/[2(2-3 \gamma)], \alpha_3=(\gamma \mp \Gamma)/2, \beta_3=1-(3 \gamma-4/3 \mp \Gamma)/[2 \gamma(\gamma \mp \Gamma)]$, and $\Gamma=[\left(-12 \gamma^{3}+45 \gamma^{2}-48 \gamma+16\right)/(-12\gamma+9)]^{1/2}$ with $\gamma$ being an arbitrary number. (d) Fourth order. The parameters are $\alpha_1=\alpha_4=\beta_1/2,\alpha_2=\alpha_3=(1-\beta_1)/2$, and $\beta_1=\beta_3=1/(2-\sqrt[3]{2}),\beta_2=-\sqrt[3]{2}\beta_1$.} \label{trotter}
 \end{figure}
 
\section{Framework}\label{framework}
We consider an $n$-qubit quantum system which is described by the system Hamiltonian $H_S$. {In order to realize a desired quantum state or quantum operation, we apply the time-dependent controls along $x$ and $y$ directions, i.e., $u(t)=(u_x^j(t),u_y^j(t))$, where $j=1,2,...,n$ and $t\in[0,T]$ with $T$ being the total time length of the pulse.}
 As such, the control Hamiltonian is written as
\begin{equation}
	H_C(t) =\sum_{j=1}^n \left[ u_x^j(t) \sigma_x^j+ u_y^j(t) \sigma_y^j \right],
\end{equation}
where $\sigma_x^j$ and $\sigma_y^j$ are the Pauli operators for the $j$th qubit. For a given time $T$, the time evolution operator of this coherently controlled quantum system is
\begin{align}\label{Eq4}
	U(T)=\mathcal{T} \exp \left\{ -i\int_0^T \left[H_S+H_C(t) \right] dt\right\},
\end{align}
with $\mathcal{T}$ being the Dyson time-ordering operator. Generally, one is not able to directly evaluate this time-dependent exponential integral in an analytical way.  The routine is to break down the continuously varying Hamiltonian into a discrete sequence. More precisely, the total evolution time $T$ is discretized into $M$ equal steps under the constraint that each duration $ \tau=T/M$ is small enough, i.e., $ \tau \ll \|H_S+H_C\|^{-1} $. As such, the control amplitude during each time duration $\tau$ can be regarded as constant, i.e., $u=(u_x^j[m],u_y^j[m])$, where $m$ runs from $1$ to $M$. After discretization, the control Hamiltonian at the $m$th step is $H_C[m] = \sum_{j=1}^{n} \left(u_{x}^{j}[m] \sigma_{x}^{j}+u_{y}^{j}[m] \sigma_{y}^{j}\right)$. {Denoting} the time evolution operator of the $m$th step as $U^{m}=e^{ -i \tau\left(H_{S}+ H_C[m] \right)}$, then the total evolution operator is $U_0(T)= U^M \cdots U^1$. {According to} the specific control target, one  defines a suitable control performance function, such as state fidelity or gate fidelity, and then various optimization algorithms can be employed to search the optimal control pulse.

A critical difficulty that arises during the iterative optimization process is that evaluating the time evolution operators, which are matrix exponentials,  may have exponential scaling of time and memory cost with respect to the number of qubits $n$. {A similar difficulty that exists in quantum simulation is realizing the complex time evolution on real physical platforms, which can be greatly solved by the well-known Trotter decompositions
\cite{suzuki1976generalized,suzuki1985decomposition,RevModPhys.86.153}. It inspires us to use the Trotter decompositions to mitigate the issue of fully computing matrix exponentials in the iterative optimization process.
Concretely,} for small enough $\tau$, we can express the $m$th step evolution operator as follows:
\begin{equation}
	e^{-i(H_S+H_C[m])\tau} = \prod_{k} e^{-i \alpha_k H_S  \tau} e^{-i \beta_k H_C[m]  \tau}  + O(\tau^{l+1}),
\end{equation}
where the parameters $\left\{ \alpha_k \right\}$ and$\left\{\beta_k\right\}$ are suitably chosen such that the right-hand side of the equation, referred to as the $l$th-order approximant,  yields an error term of order $O(\tau^{l+1})$.
{Rigorous error analysis of the Trotter decompositions should resort to the recent work Ref. \cite{PhysRevX.11.011020}.
Generally speaking, the parameters $\left\{ \alpha_k \right\}$ and$\left\{\beta_k\right\}$ can be designed symmetrically  \cite{suzuki1992sym,JWS92,HNS05} or asymmetrically \cite{ruth1983canonical,suzuki1992general,PhysRevA.102.010601}. The asymmetric scheme makes it easier to realize higher-order decompositions with rational parameters, yet requires more splitting terms compared with the symmetric scheme for the approximant of the same order \cite{suzuki1992general}. From the perspective of practicality,   we thus choose the symmetric Trotter decompositions to approximate  the time evolution operators. Specifically,}
with the simplest first-order decomposition, the system evolution operator can be approximated by
\begin{equation}	\label{eq1}
	U_1(T)  \approx  \prod_{m=1}^M e^{-i H_S \tau }e^{-i H_C[m] \tau }.
\end{equation}
 Despite its simplicity, the resultant error can become significant as the control parameters grow. To tackle this problem, we propose to use the higher-order Trotter decompositions. For the widely known second-order decomposition, the total system evolution can be approximately calculated by
\begin{equation}	\label{eq2}
	U_2(T)   \approx  \prod_{m=1}^M e^{-i H_S \tau /2}e^{-i H_C[m] \tau }e^{-i H_S \tau /2}.
\end{equation}
Similarly, the approximate system evolution operator for the third-order decomposition is
\begin{align}	\label{eq3}
U_3(T)     
	 \approx  {} & \prod_{m=1}^M e^{-i \alpha_1 H_S \tau}e^{-i \beta_1 H_C[m] \tau } e^{-i \alpha_2 H_S \tau}  \\
	&  e^{-i \beta_2 H_C[m] \tau } e^{-i \alpha_3 H_S \tau}e^{-i \beta_3 H_C[m] \tau }, \nonumber
\end{align}
and for the fourth-order decomposition is
\begin{align}	\label{eq4}
	U_4(T)     
	  \approx {} & \prod_{m=1}^M e^{-i \alpha_1 H_S \tau}e^{-i \beta_1 H_C[m] \tau } e^{-i \alpha_2 H_S \tau}  \\
	 &  e^{-i \beta_2 H_C[m] \tau } e^{-i \alpha_3 H_S \tau}e^{-i \beta_3 H_C[m] \tau } e^{-i \alpha_4 H_S \tau}. \nonumber
\end{align}
See Fig. \ref{trotter} for the illustration of the first-order to the fourth-order Trotter decompositions and the corresponding parameters $\left\{ \alpha_k \right\}$ and$\left\{\beta_k\right\}$. In this way, we split the system Hamiltonian and the control Hamiltonian into separate evolution operators. As the system evolution segments remain unchanged during the optimization, they only need to be calculated once. Additionally,  the control evolution segments are actually single-qubit operations; though renewed iteratively, they are easy to compute and store. 
We thus anticipate that this strategy can reduce the resources needed to compute the complex time evolution operator during the search of optimal control fields.

 {To figure out in which situations our strategy would likely be efficient, we provide rough analysis of the computational complexity. Generally, matrix exponential and matrix multiplication of full matrices both have the computational complexity of $O(N^3)$ with $N=2^n$ \cite{MCV03,stothers2010complexity}. Our strategy breaks the full time evolution operator into several  sparse matrix exponentials, thus the computational complexity of each  matrix exponential may reduce to $O(N^2)$ \cite{MCV03,stothers2010complexity}. However,  the increased multiplication terms will certainly decrease the computing speed as the matrix multiplication is also resource consuming. Overall, our strategy can only get modest speed gains up to several times, and is mostly favorable for large quantum systems. }

\section{Applications} \label{application}
 As a demonstration, we first test the proposed strategy with the random Ising model \cite{FAZ08} under transverse controls. The system Hamiltonian is set to be $H_S= \sum_{j=1}^{n-1}r_1^j \sigma_z^j \sigma_z^{j+1}$, where $r^1_j$ is the coefficient randomly chosen from the range $[0,1]$. The control Hamiltonian is expressed as $H_C[m]=\sum_{j=1}^n (r_2^j[m] \sigma_x^j + r_3^j[m] \sigma_y^j)$ with $r_2^j[m]$ and $r_3^j[m]$ being random values in the range $[-1,1]$. We record the computer run time required to simulate the system evolution with and without using the Trotterization technique of different orders   given by Eqs. (\ref{eq1})-(\ref{eq4}). Meanwhile,  we record the infidelity between the evolution operators simulated by the direct exponential operation ($U_0$) and the Trotter decompositions ($U_l$), i.e.,  $1-|\text{Tr}(U_l U_0^\dag)|^2/4^n$. The corresponding results are shown in Fig. \ref{fastexp}.

 \begin{figure}
\centering
\includegraphics[width=0.49\textwidth]{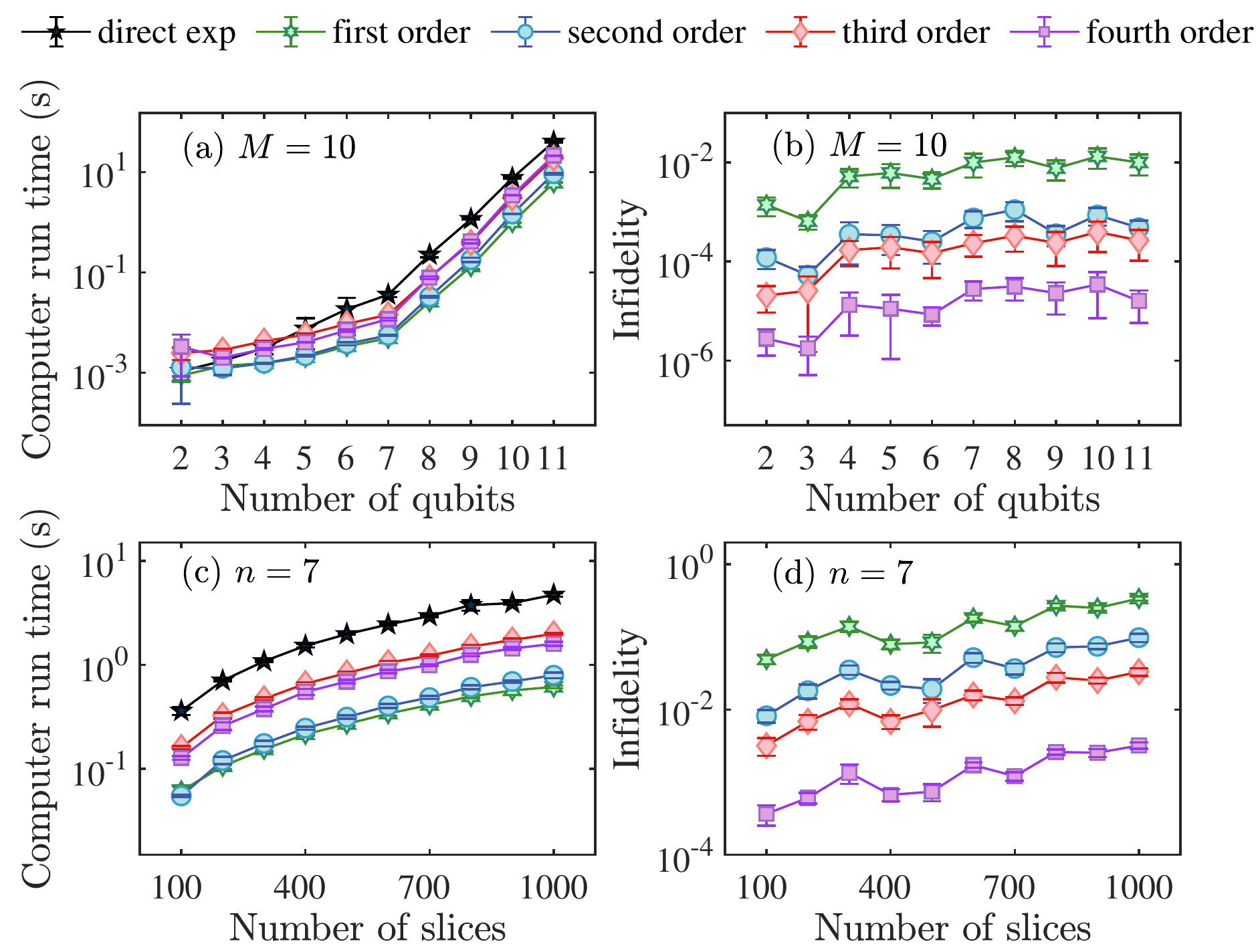}
\caption{Computer run time and infidelity ($1-$fidelity) for simulating the system evolution {of} the controlled random Ising model.
(a)-(b) For each qubit number $n$, the simulations are repeated ten times with different orders of the Trotterization to display the statistical errors, where $\tau=0.1,M=10$. (c)-(d) For the settled qubit number $n=7$ and $\tau=0.1$, the simulations for each sliced number $M$ are repeated ten times with different orders of the Trotterization to display the statistical errors.
}\label{fastexp}
 \end{figure}

 For the case of varying the qubit number $n$ shown in Fig. \ref{fastexp}(a), it is clear that using the first-order or the second-order Trotterization can greatly save the computing time, up to $50\%$ for $n=$ 2-4, and around $70$ to $90\%$ for $n=$ 5-11, while using the third-order or the fourth-order Trotterization only saves the computer run time significantly for the cases $n=$ 5-11, around $45$ to $65\%$. Nevertheless, as depicted in Fig. \ref{fastexp}(b), the lower-order Trotterizations introduce more computational errors than the higher-order Trotterizations, thus the trade-off between   accuracy and computing speed should be taken into consideration, which we will explore later. It is  worth mentioning that, owing to the memory and speed limit of the personal computer, we only simulate up to $11$ qubits with $M=10$; further simulations towards larger systems with more sliced numbers should resort to powerful supercomputers. However, in many practical situations, reliable dynamical simulations often need to discretize the time evolution into hundreds of segments, thus it is necessary to investigate the behaviors of the proposed and the conventional strategy with varied number of slices. Here, we take $n=7$ as an example to show their behaviors in Figs. \ref{fastexp}(c) and \ref{fastexp}(d). One can find that as $M$ grows, the computing time of simulating the system evolution using the Trotterizations uniformly decreases, more than $80\%$ with the first-order or the second-order Trotterization, and more than $50\%$ with the third-order or the fourth-order Trotterization. Additionally, the computational errors gradually increase as $M$ grows, which should be carefully considered in specific situations.
  
The above demonstration, though simple and direct, clearly reveals the effectiveness of the Trotterization technique on reducing the computational resources. In the following, we consider two realistic applications of searching optimal control fields using the proposed strategy.

\subsection{Improved GRAPE for pulse optimization} 
The GRAPE algorithm \cite{KNR05}, which exploits the gradient information of an objective function to update the control fields iteratively, is a well-known optimization algorithm to tackle quantum state engineering and quantum gate preparation problems. First developed for designing  nuclear magnetic resonance (NMR) pulses, it is also widely used in electron-spin resonance \cite{ZYR11}, nitrogen-vacancy centers in diamond \cite{WGW14,DFB14}, superconducting circuits \cite{MFG09,EDJ13}, ion traps \cite{NVH09,SPB11}, cold atoms \cite{SKG18}, etc. 
Normally, we first formulate an objective function $f(u)$ to assess the candidate control fields. GRAPE then finds local extrema of control solutions by taking steps along the gradient direction, i.e., $u^{(p+1)}=u^{(p)}+\lambda^{(p)}g^{(p)}$, where $p$ represents the iteration number, $\lambda^{(p)}$ is an appropriate step length and the gradient $g^{(p)}=(g_x^j[m]^{(p)},g_y^j[m]^{(p)})$ with $g_\gamma^j[m]^{(p)}=\partial f ^{(p)}/\partial u_\gamma^j[m]^{(p)}$ for $\gamma=x,y$.
{To accelerate the convergence speed, we apply the second-order quasi-Newton method, which approximates the actual Hessian with the gradient information computed from previous steps \cite{PhysRevA.84.022305,de2011second}. More precisely, this is achieved by using the Limited-memory Broyden–Fletcher–Goldfarb–Shanno (LBFGS) method in Matlab, which further optimizes the optimization procedure with limited memory \cite{liu1989limited}.}
Clearly, in  the process of calculating the objective function $f$ {and its approximate Hessian}, a considerable number of matrix exponentials of the time evolution segments need to be done, which would cost an  exponential amount of   computer run time. Furthermore, these time evolution segments must be re-evaluated in each iteration for the renewed control parameters, making the algorithm quickly intractable for even modest-sized quantum optimal control problems. Thus it is desirable to develop more efficient strategies to simulate the system evolution here. {It is worth noting that  Ref. \cite{BJ18} has concisely discussed how to use the first- and the second-order Trotterizations to improve the basic GRAPE algorithm. Our applications here explore in detail the performance of the high-order Trotterizations in pulse optimization with the second-order GRAPE algorithm.  }
  
  \begin{figure}
\centering
\includegraphics[width=0.48\textwidth]{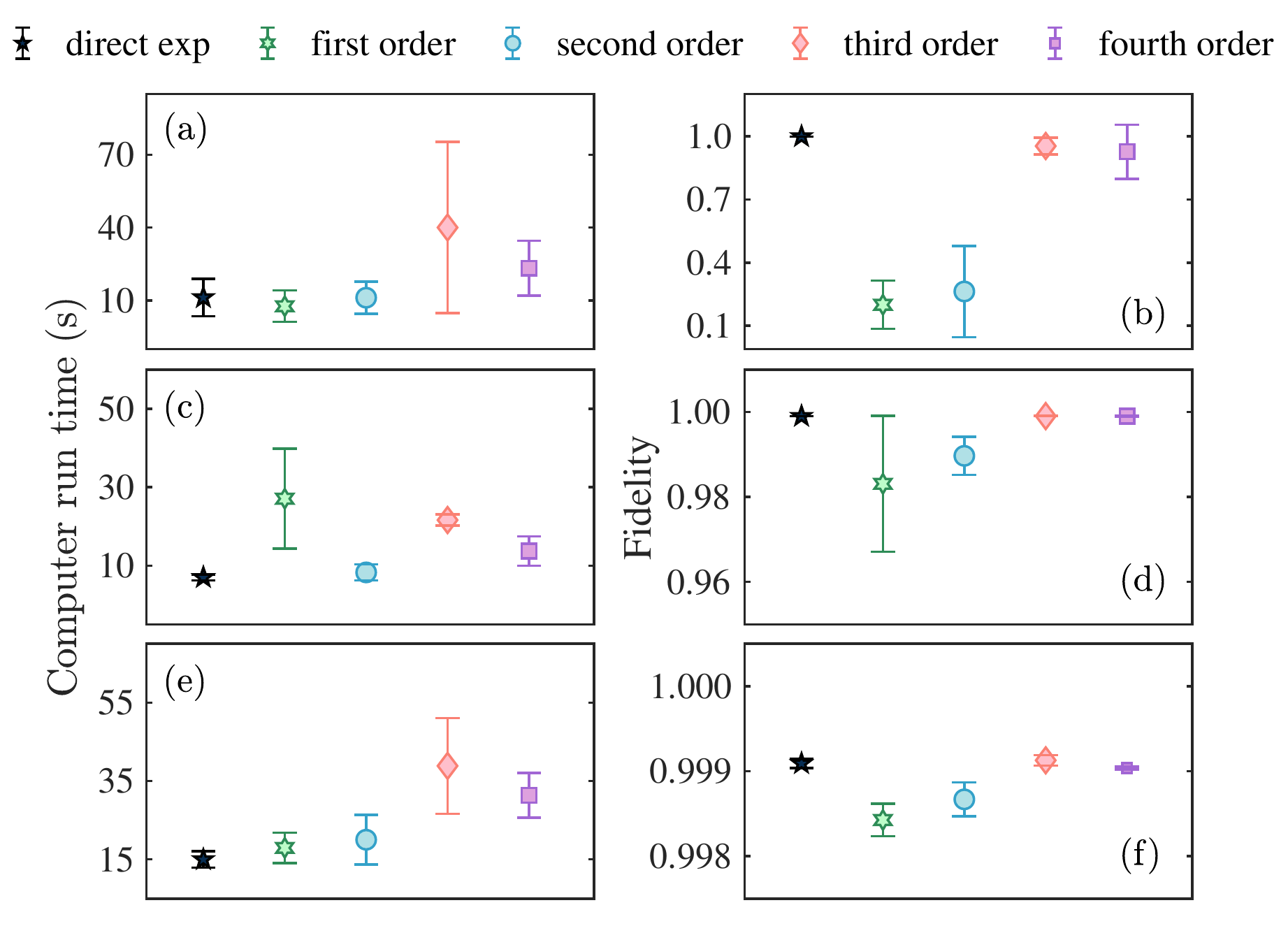}
\caption{The GRAPE algorithm with the Trotterizations for searching optimal pulses to prepare the four-qubit GHZ state. The total evolution time $T=5~$ms, which is divided into different number of slices: (a)-(b) $M=200$; (c)-(d) $M=500$; (e)-(f) $M=1000$. The optimization is terminated when the estimated fidelity $f_l>0.999$ or the iteration number $p>1000$. The computer run time and the final true fidelity $f$ are recorded. The optimizations are all repeated five times to display the statistical errors.   }\label{4qubit}
 \end{figure}

 To give a concrete example, consider the control optimization problem of preparing the known  Greenberger-Horne-Zeilinger (GHZ) state \cite{PLS18} in a four-qubit NMR system. The natural Hamiltonian can be expressed as $H_S=-\sum_{i=1}^4 {\omega_{i}}\sigma^i_z/2+\sum_{i<j,=1}^4 {\pi J_{ij}}\sigma^i_z \sigma^j_z/2$, where $\omega_i$ represents the Larmor precession frequency for the $i$th spin and $J_{ij}$ is the $J$-coupling constant between the $i$th and the $j$th spin. The strengths of the Larmor frequencies and the $J$ couplings can be found in Ref. \cite{ZHJ20}. The four-qubit GHZ state takes the form $|\psi_\text{GHZ}\rangle = (|0\rangle ^{\otimes 4} + |1\rangle ^{\otimes 4})/\sqrt{2}$.  The goal is to find an optimal control pulse $u_{\text{opt}}$ that can steer the system from $|\psi_0\rangle=|0\rangle^{\otimes 4} $ to $|\psi_\text{GHZ}\rangle$, with maximizing the state fidelity defined as $f_l= \left|\langle \psi_\text{GHZ} | U_l |\psi_0\rangle \right|^2$. Combining the Trotterization technique with GRAPE gives a way to find the optimal control fields; a similar work is found in Ref. \cite{BJ18}. However, the Trotterization technique may introduce significant errors in simulating the system evolution, thus we check the true fidelity of the discovered optimal controls by $f=\left|\langle \psi_\text{GHZ} | U_0(u_{\text{opt}}) |\psi_0\rangle \right|^2$. The computer run time and the true fidelity using different orders of the Trotterization are shown in Fig. \ref{4qubit}. We fix the total evolution time as $T=5~$ ms with different number of slices in the optimizations. For the first-order and the second-order Trotterization, if $M=200$, it can slightly save the computing time, yet introducing very large computational errors [see Figs. \ref{4qubit}(a) and \ref{4qubit}(b)]. Increasing the sliced number $M$ can improve the true fidelity [Figs. \ref{4qubit}(d) and  \ref{4qubit}(f)], but cost around {18-286\%} more computing time [Figs. \ref{4qubit}(c) and  \ref{4qubit}(e)]. For the third-order and the fourth-order Trotterization, though they always induce very high true fidelity beyond 0.99, and around {85-264\%} more computing time is needed. In total, the Trotterization technique does not provide speedup for the GRAPE algorithm in a small-scale system, which is consistent with the results shown in Fig. \ref{fastexp}. {The reason behind this is that matrix exponential and matrix multiplication both generally scale as $O(N^3)$ with $N=2^n$. Though the splitting terms using the Trotterization may scale as $O(N^2)$, the increased multiplications will reduce the computing speed. Thus for a small-scale system, the Trotterization based GRAPE algorithm may fail to achieve speed gains.}
  
\begin{figure}
\centering
\includegraphics[width=0.48\textwidth]{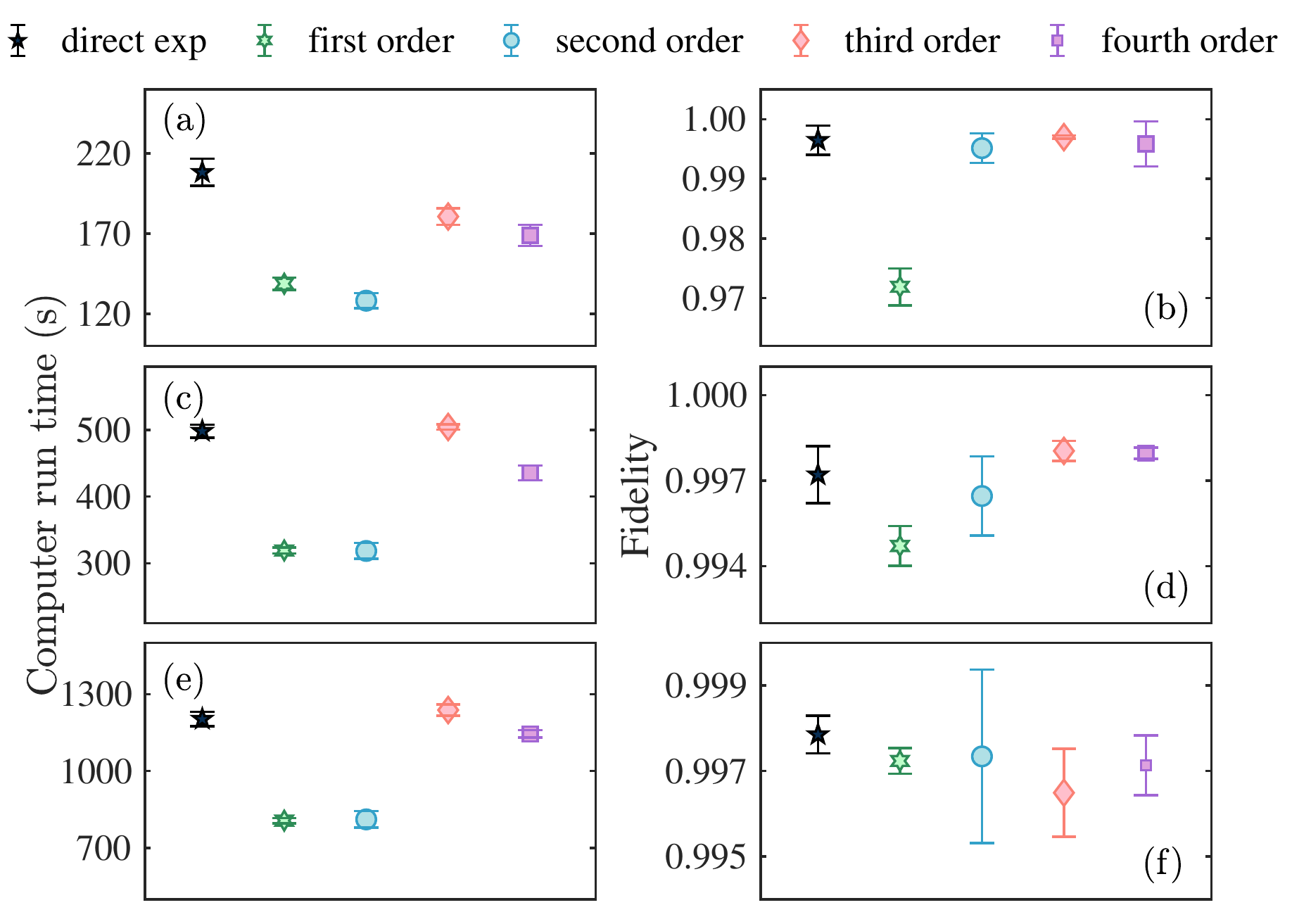}
\caption{The GRAPE algorithm with the Trotterizations for searching optimal pulses to realize single-qubit rotation in a seven-qubit system. The total evolution time $T=1~$ms, which is divided into different number of slices: (a)-(b) $M=100$; (c)-(d) $M=250$; (e)-(f) $M=500$. We stop the optimization when the estimated fidelity $f_l>0.999$ or the iteration number $p$ exceeds $500$, and record the computer run time and the final true fidelity $f$ for each case. The optimizations are all repeated five times to display the statistical errors. }\label{7qubit}
 \end{figure}

 Now we turn to consider the problem of finding optimal pulses for achieving single-qubit rotation in a seven-qubit NMR system. The natural Hamiltonian is $H_S=-\sum_{i=1}^7 {\omega_{i}}\sigma^i_z/2+\sum_{i<j,=1}^7 {\pi J_{ij}}\sigma^i_z \sigma^j_z/2$, 
% where $\omega_i$ represent the Larmor precession frequency for the $i$th spin and $J_{ij}$ is the $J$-coupling constant between the $i$th and the $j$th spin. 
and the strengths of the Larmor frequencies and the $J$ couplings can be found in Ref. \cite{LYP17}. Our goal is to search out an optimal control pulse $u_{\text{opt}}$ that can steer the system to achieve the target single-qubit operation $U_t=\exp(-i \pi \sigma^2_x/4)$, with maximizing the gate fidelity $f_l=|\text{Tr}(U_l U_t^\dag)|^2/4^7$. Similarly, we combine the Trotterization technique with the GRAPE algorithm to accomplish this optimization task, and we check the final true fidelity of the searched optimal controls with $f=|\text{Tr}(U_0(u_{\text{opt}}) U_t^\dag)|^2/4^7$. We fix the total evolution time as $T=1~$ ms with different number of slices in the optimizations. For the case of $M=100$ shown in Figs. \ref{7qubit}(a) and \ref{7qubit}(b), using the first-order to the fourth-order Trotterization can save the computing time about {33, 38, 13, and 19\%}, respectively. However, the final true gate fidelities using the first-order and the second-order Trotterization are roughly smaller than 0.99, which is often below the fault-tolerant   threshold \cite{CET17}. Increasing $M$ to 250, as shown in Figs. \ref{7qubit}(c) and \ref{7qubit}(d), the final gate fidelities are all beyond {0.994}, and the first-order to the fourth-order Trotterization can save the computing time about {36, 36, -1, and 12\%}, respectively. For much larger sliced number $M=500$ shown in Figs. \ref{7qubit}(e) and \ref{7qubit}(f), their true fidelities are all beyond {0.995} and the corresponding reduced computing time is about {33, 32, -3,  and 5\%}, respectively. These results indicate that using the first-order and the second-order Trotterization can always significantly save the computing time, but only achieve sufficient high fidelity when the duration time $\tau=T/M$ is small enough. Meanwhile, using the third-order and the fourth-order Trotterization can achieve very high fidelity even when $\tau$ is relatively large, but may not save that much computing time. 
 In total, these results indicate that our proposed strategy is mostly favorable for a relatively large system, and it can possibly save much more computer run time when searching optimal pulses for intermediate-scale optimization tasks. 

{To further exploit the potential of the proposed strategy, we attempt to use hybrid Trotterizations to improve the performance of the GRAPE. Precisely, we apply the low-order Trotterization in the early stages of GRAPE and the switch to the high-order Trotterization in the final stages. To test this hybrid strategy, the specific problem we choose  is the same as above, namely, finding optimal pulses for realizing single-qubit rotation in a seven-qubit NMR system.  In the simulations, we use the first-order or the second-order Trotterization during the former 400 iterations and then switch to the third-order or the fourth-order Trotterization during the latter 100 iterations. The simulation results are shown in Fig. \ref{hybrid}. Compared with the first-order Trotterization, it is clear that the hybrid first- to third-order and first- to fourth-order Trotterization significantly increase the fidelity from 0.972 to 0.993, nearly without costing more computer run time, while, compared with second-order Trotterization, the hybrid second- to third-order and second- to fourth-order Trotterization can slightly increase the fidelity from 0.995 to 0.997, but cost 30-40\% more computer run time. These results reveal that with proper hybrid Trotterizations, the performance of the GRAPE algorithm can be further improved. 
}
 \begin{figure}
\centering
\includegraphics[width=0.46\textwidth]{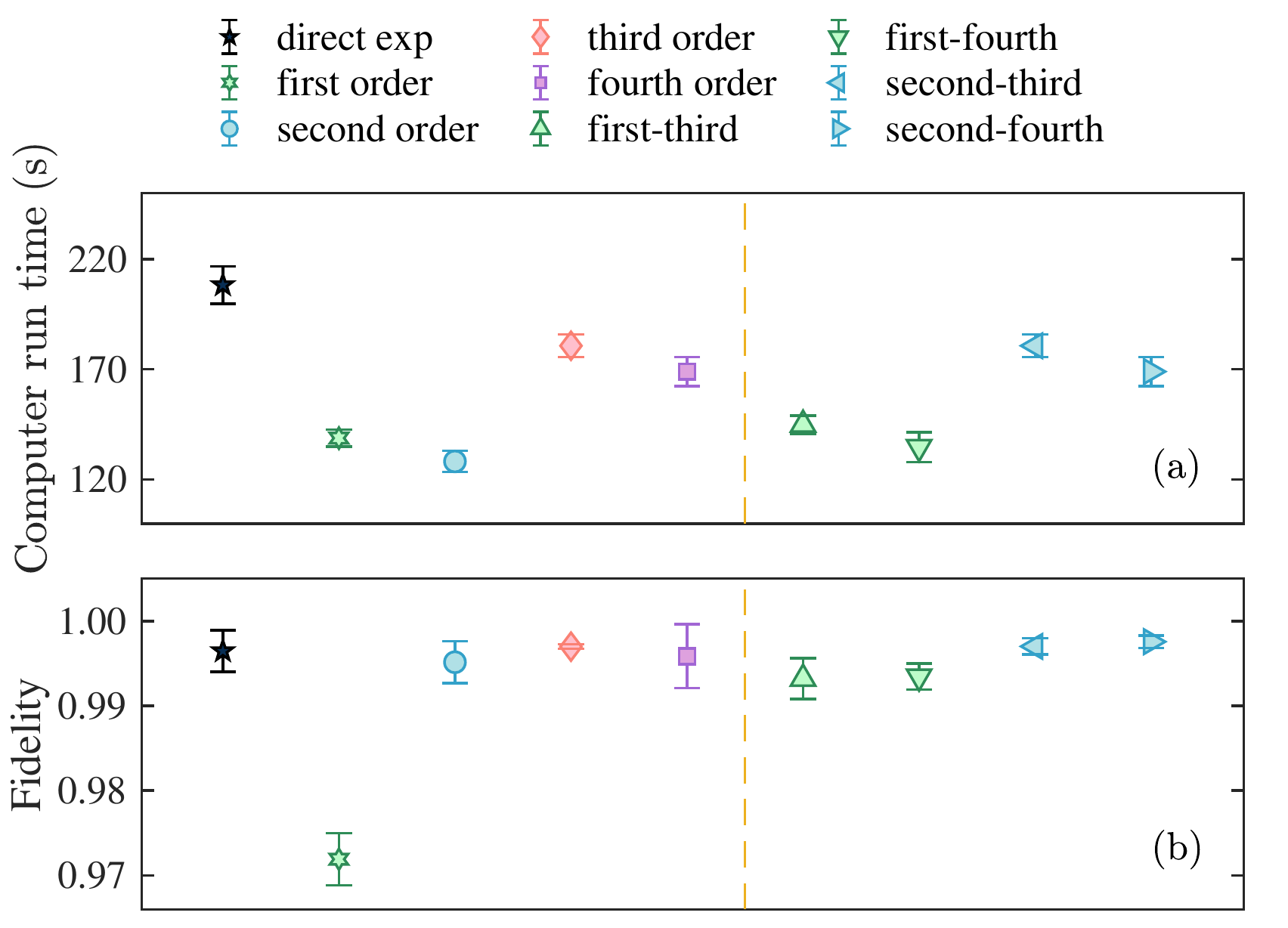}
\caption{{The GRAPE algorithm with different Trotterization strategies for searching optimal pulses to realize single-qubit rotation in a seven-qubit system, where $T=1$ and $M=100$. (a) and (b) display the computer run time and the fidelity for different Trotterization strategies, respectively. The panels on the right of the yellow dashed lines represent the results using the hybrid Trotterization strategies. The stopping condition is that the estimated fidelity $f_l>0.999$ or the iteration number $p>500$.
 }}\label{hybrid}
 \end{figure}

 \subsection{Improved VQA for ground-state energy solving}
 
 \begin{figure*}
\centering
\includegraphics[width=0.75\textwidth]{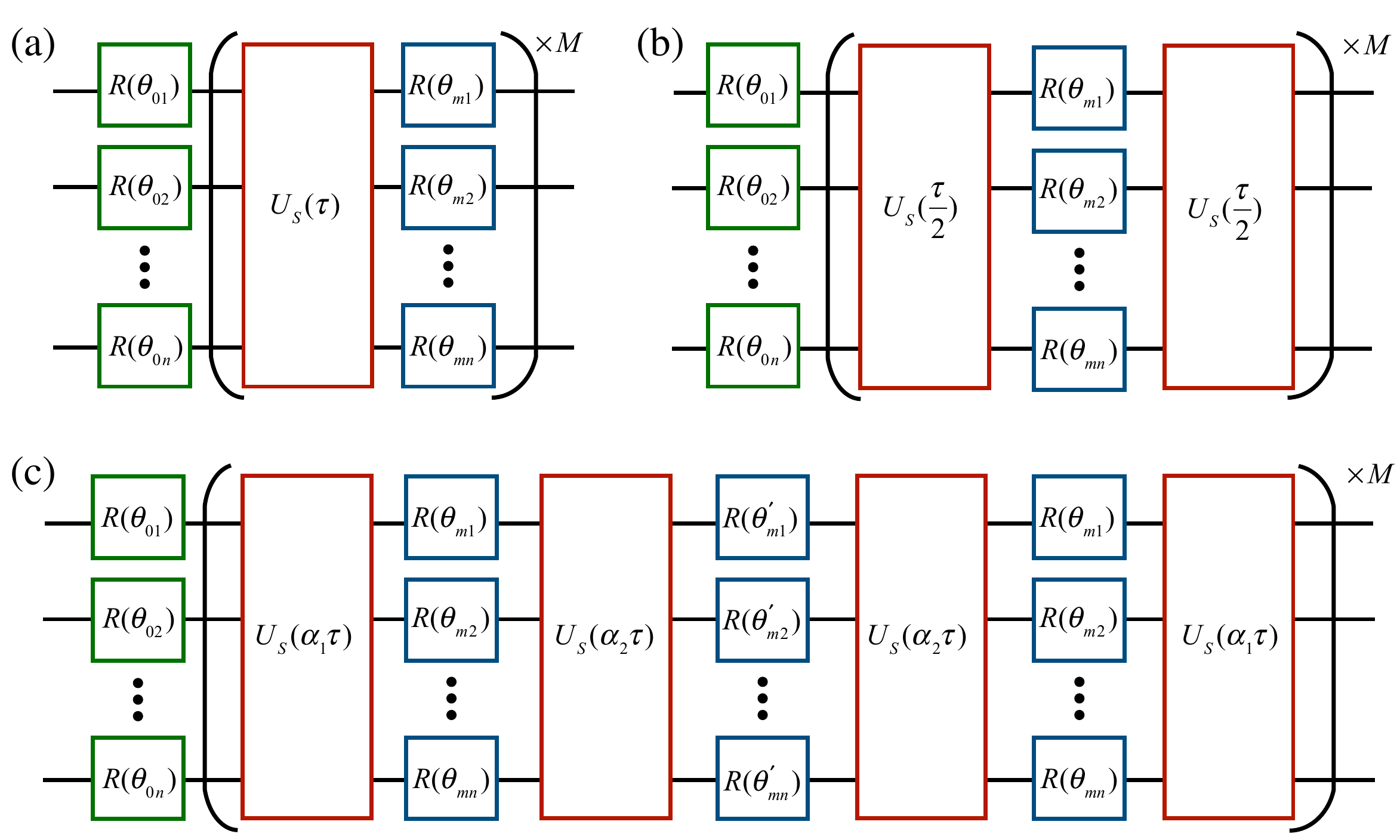}
\caption{Efficient Ans\"atze inspired by the symmetric Trotterization for VQA. (a) Illustration of VQA based on the first-order Trotterization. This structure is also refereed to as a hardware-efficient Ansatz in Ref. \cite{KAM17}. (b) and (c) VQA based on the second-order and the symmetric fourth-order Trotterization, respectively, where $\alpha_1=s/2,\alpha_2=(1-s)/2, \theta_{jm}^\prime=-\sqrt[3]{2} \theta_{jm}$ with $s=1/(2-\sqrt[3]{2}), j=1,2,...,n$. In all the figures, $U_S(\tau)$ is the free evolution operator during the period $\tau$, which is governed by the available system Hamiltonian $H_S$, i.e., $U_S(\tau)=\exp(-i  H_{S} \tau)$. Each single-qubit rotation is represented by $R(\theta_{mj})$ with the angle $\theta_{mj}$, and the rotation axis is typically chosen as $x$ axis and $y$ axis.  }\label{ansatz}
 \end{figure*}

 The VQA \cite{MNB18,CMA20}, which functions by minimizing certain cost functions via the variation of limited gate parameters with low-depth circuits, has found tremendous applications in quantum chemistry, quantum simulation and machine learning \cite{CMA20}.  Usually, the VQA first parametrizes a quantum circuit $U(\bm \theta)$ with multiple layers of building blocks (called Ansatz), with each block involving single-qubit rotations and available non-local gates. Suitable optimization algorithms are then applied to tune the parameters $\bm \theta$ for minimizing the target cost function. The structure of an Ansatz is generally designed according to the optimization task at hand, but it can also be formalized when no relevant information is readily known. A notable example hardware-efficient Ansatz \cite{KAM17} (see Fig. \ref{ansatz}(a)), which utilizes available entangling resources in a given physical device, is prominent for problem-agnostic situations. 
   
  Here, we find that the hardware-efficient Ansatz actually has a strong relationship with the sliced time evolution simulated by the first-order Trotterization. To be specific, the $m$th sliced evolution can be  equivalently expressed as
\begin{equation}
U_m  \approx  e^{-i H_S \tau}  e^{-i H_C[m] \tau}    
	 \equiv  e^{-i H_S \tau} \cdot \bigotimes_{j=1}^n R( {\theta}_{mj}) ,
\end{equation}
where the control Hamiltonian part is the product of generic single-qubit rotations represented by 
\begin{equation}
	R( \theta_{mj}) =R_x(\theta_{mj}^1) R_y(\theta_{mj}^2) R_x(\theta_{mj}^3),
\end{equation}
and $R_\gamma( \theta_{mj}^w) = \exp (-i \theta_{mj}^w\sigma_\gamma^j /2 ), \gamma=x,y;w=1,2,3$.	
Following similar mapping rules, we propose  Ans\"atze based on the second-order and the fourth-order symmetric Trotterization \cite{JWS92,KHH94,HNS05}, as shown in Figs. \ref{ansatz}(b) and \ref{ansatz}(c). For example, for the second-order Trotterization, the $m$th sliced evolution operator can be written as
\begin{align}
	U_m & \approx  e^{-i H_S \tau/2}  e^{-i H_C[m] \tau}  e^{-i H_S \tau/2}  \nonumber \\
	& =  e^{-i H_S \tau/2} \cdot \bigotimes_{j=1}^n R( {\theta}_{mj}) \cdot   e^{-i H_S \tau/2}.
	\label{second}
\end{align}
As such, the whole evolution becomes a parametrized quantum circuit as shown in Fig. \ref{ansatz}(b). A similar {second-order Trotterization based} Ansatz has been used in Ref. \cite{WDH15}, and recent works \cite{CAD20,MAB21} also provide insights into informing the VQA with the quantum optimal control perspective. As we have analyzed that the higher-order Trotterizations can help improve the efficiency of simulating system evolution, we thus expect that the proposed new Ans\"atze function in improving the VQA. Additionally, it should be noted that for all the proposed Ans\"atze,  the overall free evolution time in each block is $\tau$ , and the total number of parameters is $3(M+1)n$. This indicates that the higher-order Trotterization inspired Ans\"atze will not introduce extra computational and operational burdens.

\begin{figure}
\centering
\includegraphics[width=0.48\textwidth,height=0.34\textwidth]{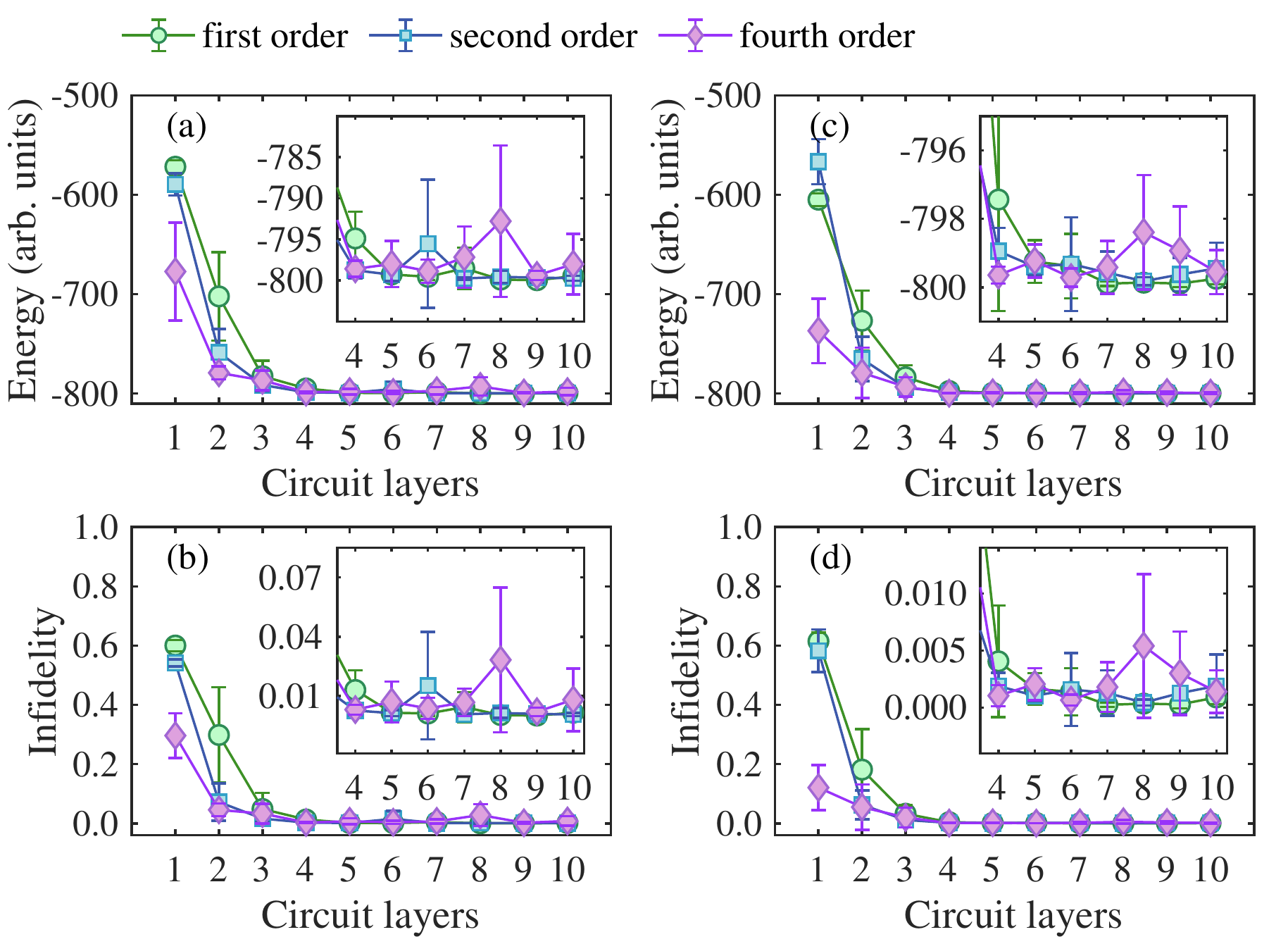}
\caption{The Trotterization inspired VQA for solving the ground-state energy in the four-qubit Heisenberg model on a square lattice. Minimal energy found using VQA of different circuit layers when (a) $\tau=0.005~$s; (c) $\tau=0.01~$s. Corresponding state infidelity between the searched optimal ground state and the theoretical ground state when (b) $\tau=0.005~$s; (d) $\tau=0.01~$s. 
The optimization stops when the maximal iteration number exceeds 1000, and for each circuit layer the optimizations are repeated five times to display the statistical errors.  }\label{HM}
 \end{figure}

As a demonstration, we consider the problem of solving the ground-state energy in a four-qubit Heisenberg spin model on a square lattice \cite{KAM17}. This typical model has been widely explored in quantum areas, from thermodynamics and statistics to communication and computation \cite{BMC11, ZDG09, DGL12}, which can be realized by various physical systems \cite{SC1, SC2}, such as ultracold atoms, trapped ions, and nuclear magnetic resonance. This model can be described by
\begin{equation}
	 H_P =J \sum_{\langle i j\rangle}\left(\sigma_x^{i} \sigma_x^{j}+\sigma_y^{i} \sigma_y^{j}+\sigma_z^{i} \sigma_z^{j}\right)+B_z \sum_{i} \sigma_z^{i},
\end{equation}
where $J$ is the nearest-neighbor interaction strength,  and $B_{z}$ is the longitudinal static field. With loss of generality, we set $J=100,B_z=100$.

To tackle this eigensolver problem, we use a four-spin NMR processor to generate candidate states from $\rho_0=|0\rangle^{\otimes 4}$ with the Trotterization inspired VQA. We then iteratively update the parameters to minimize the energy $f=\text{Tr}(H_P U(\bm \theta) \rho_0 U(\bm \theta)^\dag)$ using the improved Nelder-Mead algorithm \cite{PNM11}. The natural Hamiltonian of this NMR processor is $H_S=-\sum_{i=1}^4 {\omega_{i}}\sigma^i_z/2+\sum_{i<j,=1}^4 {\pi J_{ij}}\sigma^i_z \sigma^j_z/2$, where $\omega_i$ represents the Larmor precession frequency for the $i$th spin and $J_{ij}$ is the $J$-coupling constant between the $i$th and the $j$th spin. The strengths of the Larmor frequencies and the $J$ couplings can be found in Ref. \cite{ZHJ20}. With different circuit layers, we show the numerical energy optimization results in Figs. \ref{HM}(a) and \ref{HM}(c). By defining the state fidelity between the searched ground state and the theoretical ground state $F=\text{Tr}(\rho_{th} U(\bm \theta) \rho_0 U(\bm \theta)^\dag)$, we also display the corresponding state infidelity ($1-F$) in Figs. \ref{HM}(b)  and \ref{HM}(d).
 These numerical results indicate that for small number of circuit layers $M=$ 1-3, the second-order and the fourth-order Trotterization inspired Ans\"atze perform much better than the hardware-efficient Ansatz.
  When increasing the circuit layers ($M>3$), the second-order Trotterization and the fourth-order Trotterization inspired Ans\"atze will have comparable performance with the hardware-efficient Ansatz.
 This is reasonable because for a limited number of circuit layers $M$, the higher-order Trotter decompositions possess higher precision to simulate the system evolution than the conventional hardware-efficient Ansatz. However, when $M$ is sufficiently large, the computational resources will be enough for all orders of the Trotter decompositions to achieve accurate results.
  This indicates that the proposed Ans\"atze are favorable for the highly short-depth VQA to find approximate solutions.
 
\section{Conclusion and discussion} \label{discussion}
In the design of precise quantum control, simulating the system time evolution is the most resource-consuming part. To mitigate this issue, we combine the Trotter decompositions to propose a practical strategy suitable for various control optimization problems. The demonstrations with the GRAPE algorithm up to seven qubits show the effectiveness of our strategy in reducing the computing time. We expect that this strategy can be explored in an intermediate-scale system in the future, probably combining parallelization technique \cite{GTS06} or special matrix representation method \cite{HHH10}.  
Several new Ans\"atze inspired by the Trotter decompositions are also presented for improving the performance of the VQA, showing the advantages of finding approximate solutions with highly low-depth circuits. Actually, this is favorable for realistic applications, as current quantum devices often process many noisy qubits with a limited coherence time \cite{LTD10}. Furthermore, future investigations can explore the possibility of implementing the proposed strategy in more complex optimization and simulation tasks.

\section*{Acknowledgments} 
This work was supported by the National Natural Science Foundation of China (Grants No. 12204230, No. 1212200199, No. 11975117, No. 12075110, No. 11905099, No. 11875159, No. 11905111, No. U1801661, and No. 92065111); National Key Research and Development Program of China (Grant No. 2019YFA0308100); Guangdong Basic and Applied Basic Research Foundation (Grants No. 2019A1515011383 and No. 2021B1515020070); Guangdong Provincial Key Laboratory (Grant No. 2019B121203002); Guangdong International Collaboration Program (Grant No. 2020A0505100001); Shenzhen Science and Technology Program (Grants No. RCYX20200714114522109 and No. KQTD20200820113010023); China Postdoctoral Science Foundation (Grant No. 2021M691445); Science, Technology, and Innovation Commission of Shenzhen Municipality (Grants No. ZDSYS20190902092905285, No. KQTD20190929173815000, No. JCYJ20200109140803865, and No. JCYJ20180302174036418); and Pengcheng Scholars, Guangdong Innovative and Entrepreneurial Research Team Program (Grant No. 2019ZT08C044).

\providecommand{\noopsort}[1]{}\providecommand{\singleletter}[1]{#1}%


\begin{thebibliography}{0}%
\makeatletter
\providecommand \@ifxundefined [1]{%
 \@ifx{#1\undefined}
}%
\providecommand \@ifnum [1]{%
 \ifnum #1\expandafter \@firstoftwo
 \else \expandafter \@secondoftwo
 \fi
}%
\providecommand \@ifx [1]{%
 \ifx #1\expandafter \@firstoftwo
 \else \expandafter \@secondoftwo
 \fi
}%
\providecommand \natexlab [1]{#1}%
\providecommand \enquote  [1]{``#1''}%
\providecommand \bibnamefont  [1]{#1}%
\providecommand \bibfnamefont [1]{#1}%
\providecommand \citenamefont [1]{#1}%
\providecommand \href@noop [0]{\@secondoftwo}%
\providecommand \href [0]{\begingroup \@sanitize@url \@href}%
\providecommand \@href[1]{\@@startlink{#1}\@@href}%
\providecommand \@@href[1]{\endgroup#1\@@endlink}%
\providecommand \@sanitize@url [0]{\catcode `\\12\catcode `\$12\catcode
  `\&12\catcode `\#12\catcode `\^12\catcode `\_12\catcode `\%12\relax}%
\providecommand \@@startlink[1]{}%
\providecommand \@@endlink[0]{}%
\providecommand \url  [0]{\begingroup\@sanitize@url \@url }%
\providecommand \@url [1]{\endgroup\@href {#1}{\urlprefix }}%
\providecommand \urlprefix  [0]{URL }%
\providecommand \Eprint [0]{\href }%
\providecommand \doibase [0]{https://doi.org/}%
\providecommand \selectlanguage [0]{\@gobble}%
\providecommand \bibinfo  [0]{\@secondoftwo}%
\providecommand \bibfield  [0]{\@secondoftwo}%
\providecommand \translation [1]{[#1]}%
\providecommand \BibitemOpen [0]{}%
\providecommand \bibitemStop [0]{}%
\providecommand \bibitemNoStop [0]{.\EOS\space}%
\providecommand \EOS [0]{\spacefactor3000\relax}%
\providecommand \BibitemShut  [1]{\csname bibitem#1\endcsname}%
\let\auto@bib@innerbib\@empty
%</preamble>
\end{thebibliography}%


\begin{thebibliography}{66}%
\makeatletter
\providecommand \@ifxundefined [1]{%
 \@ifx{#1\undefined}
}%
\providecommand \@ifnum [1]{%
 \ifnum #1\expandafter \@firstoftwo
 \else \expandafter \@secondoftwo
 \fi
}%
\providecommand \@ifx [1]{%
 \ifx #1\expandafter \@firstoftwo
 \else \expandafter \@secondoftwo
 \fi
}%
\providecommand \natexlab [1]{#1}%
\providecommand \enquote  [1]{``#1''}%
\providecommand \bibnamefont  [1]{#1}%
\providecommand \bibfnamefont [1]{#1}%
\providecommand \citenamefont [1]{#1}%
\providecommand \href@noop [0]{\@secondoftwo}%
\providecommand \href [0]{\begingroup \@sanitize@url \@href}%
\providecommand \@href[1]{\@@startlink{#1}\@@href}%
\providecommand \@@href[1]{\endgroup#1\@@endlink}%
\providecommand \@sanitize@url [0]{\catcode `\\12\catcode `\$12\catcode
  `\&12\catcode `\#12\catcode `\^12\catcode `\_12\catcode `\%12\relax}%
\providecommand \@@startlink[1]{}%
\providecommand \@@endlink[0]{}%
\providecommand \url  [0]{\begingroup\@sanitize@url \@url }%
\providecommand \@url [1]{\endgroup\@href {#1}{\urlprefix }}%
\providecommand \urlprefix  [0]{URL }%
\providecommand \Eprint [0]{\href }%
\providecommand \doibase [0]{https://doi.org/}%
\providecommand \selectlanguage [0]{\@gobble}%
\providecommand \bibinfo  [0]{\@secondoftwo}%
\providecommand \bibfield  [0]{\@secondoftwo}%
\providecommand \translation [1]{[#1]}%
\providecommand \BibitemOpen [0]{}%
\providecommand \bibitemStop [0]{}%
\providecommand \bibitemNoStop [0]{.\EOS\space}%
\providecommand \EOS [0]{\spacefactor3000\relax}%
\providecommand \BibitemShut  [1]{\csname bibitem#1\endcsname}%
\let\auto@bib@innerbib\@empty
%</preamble>
\bibitem [{\citenamefont {Warren}\ \emph {et~al.}(1993)\citenamefont {Warren},
  \citenamefont {Rabitz},\ and\ \citenamefont {Dahleh}}]{WWS93}%
  \BibitemOpen
  \bibfield  {author} {\bibinfo {author} {\bibfnamefont {W.~S.}\ \bibnamefont
  {Warren}}, \bibinfo {author} {\bibfnamefont {H.}~\bibnamefont {Rabitz}},\
  and\ \bibinfo {author} {\bibfnamefont {M.}~\bibnamefont {Dahleh}},\
  }\bibfield  {title} {\bibinfo {title} {Coherent Control of Quantum Dynamics:
  The Dream Is Alive},\ }\href {https://doi.org/10.1126/science.259.5101.1581}
  {\bibfield  {journal} {\bibinfo  {journal} {Science}\ }\textbf {\bibinfo
  {volume} {259}},\ \bibinfo {pages} {1581} (\bibinfo {year}
  {1993})}\BibitemShut {NoStop}%
\bibitem [{\citenamefont {Brif}\ \emph {et~al.}(2010)\citenamefont {Brif},
  \citenamefont {Chakrabarti},\ and\ \citenamefont {Rabitz}}]{BCC10}%
  \BibitemOpen
  \bibfield  {author} {\bibinfo {author} {\bibfnamefont {C.}~\bibnamefont
  {Brif}}, \bibinfo {author} {\bibfnamefont {R.}~\bibnamefont {Chakrabarti}},\
  and\ \bibinfo {author} {\bibfnamefont {H.}~\bibnamefont {Rabitz}},\
  }\bibfield  {title} {\bibinfo {title} {Control of quantum phenomena: past,
  present and future},\ }\href {https://doi.org/10.1088/1367-2630/12/7/075008}
  {\bibfield  {journal} {\bibinfo  {journal} {New J. Phys.}\ }\textbf {\bibinfo
  {volume} {12}},\ \bibinfo {pages} {075008} (\bibinfo {year}
  {2010})}\BibitemShut {NoStop}%
\bibitem [{\citenamefont {Koch}\ \emph {et~al.}(2022)\citenamefont {Koch},
  \citenamefont {Boscain}, \citenamefont {Calarco}, \citenamefont {Dirr},
  \citenamefont {Filipp}, \citenamefont {Glaser}, \citenamefont {Kosloff},
  \citenamefont {Montangero}, \citenamefont {Schulte-Herbr{\"u}ggen},
  \citenamefont {Sugny} \emph {et~al.}}]{koch2022quantum}%
  \BibitemOpen
  \bibfield  {author} {\bibinfo {author} {\bibfnamefont {C.~P.}\ \bibnamefont
  {Koch}}, \bibinfo {author} {\bibfnamefont {U.}~\bibnamefont {Boscain}},
  \bibinfo {author} {\bibfnamefont {T.}~\bibnamefont {Calarco}}, \bibinfo
  {author} {\bibfnamefont {G.}~\bibnamefont {Dirr}}, \bibinfo {author}
  {\bibfnamefont {S.}~\bibnamefont {Filipp}}, \bibinfo {author} {\bibfnamefont
  {S.~J.}\ \bibnamefont {Glaser}}, \bibinfo {author} {\bibfnamefont
  {R.}~\bibnamefont {Kosloff}}, \bibinfo {author} {\bibfnamefont
  {S.}~\bibnamefont {Montangero}}, \bibinfo {author} {\bibfnamefont
  {T.}~\bibnamefont {Schulte-Herbr{\"u}ggen}}, \bibinfo {author} {\bibfnamefont
  {D.}~\bibnamefont {Sugny}}, \emph {et~al.},\ }\bibfield  {title} {\bibinfo
  {title} {Quantum optimal control in quantum technologies. Strategic report on
  current status, visions and goals for research in europe},\ }\href
  {https://doi.org/10.1140/epjqt/s40507-022-00138-x} {\bibfield  {journal}
  {\bibinfo  {journal} {EPJ Quantum Technol.}\ }\textbf {\bibinfo {volume}
  {9}},\ \bibinfo {pages} {19} (\bibinfo {year} {2022})}\BibitemShut {NoStop}%
\bibitem [{\citenamefont {Boscain}\ \emph {et~al.}(2021)\citenamefont
  {Boscain}, \citenamefont {Sigalotti},\ and\ \citenamefont
  {Sugny}}]{PRXQuantum.2.030203}%
  \BibitemOpen
  \bibfield  {author} {\bibinfo {author} {\bibfnamefont {U.}~\bibnamefont
  {Boscain}}, \bibinfo {author} {\bibfnamefont {M.}~\bibnamefont {Sigalotti}},\
  and\ \bibinfo {author} {\bibfnamefont {D.}~\bibnamefont {Sugny}},\ }\bibfield
   {title} {\bibinfo {title} {Introduction to the Pontryagin Maximum Principle
  for Quantum Optimal Control},\ }\href
  {https://doi.org/10.1103/PRXQuantum.2.030203} {\bibfield  {journal} {\bibinfo
   {journal} {PRX Quantum}\ }\textbf {\bibinfo {volume} {2}},\ \bibinfo {pages}
  {030203} (\bibinfo {year} {2021})}\BibitemShut {NoStop}%
\bibitem [{\citenamefont {d'Alessandro}(2007)}]{DAD07}%
  \BibitemOpen
  \bibfield  {author} {\bibinfo {author} {\bibfnamefont {D.}~\bibnamefont
  {d'Alessandro}},\ }\href {https://doi.org/10.1201/9781584888833} {\emph
  {\bibinfo {title} {Introduction to Quantum Control and Dynamics}}}\ (\bibinfo
   {publisher} {CRC Press, New York},\ \bibinfo {year} {2007})\BibitemShut
  {NoStop}%
\bibitem [{\citenamefont {Khaneja}\ \emph {et~al.}(2005)\citenamefont
  {Khaneja}, \citenamefont {Reiss}, \citenamefont {Kehlet}, \citenamefont
  {Schulte-Herbr{\"u}ggen},\ and\ \citenamefont {Glaser}}]{KNR05}%
  \BibitemOpen
  \bibfield  {author} {\bibinfo {author} {\bibfnamefont {N.}~\bibnamefont
  {Khaneja}}, \bibinfo {author} {\bibfnamefont {T.}~\bibnamefont {Reiss}},
  \bibinfo {author} {\bibfnamefont {C.}~\bibnamefont {Kehlet}}, \bibinfo
  {author} {\bibfnamefont {T.}~\bibnamefont {Schulte-Herbr{\"u}ggen}},\ and\
  \bibinfo {author} {\bibfnamefont {S.~J.}\ \bibnamefont {Glaser}},\ }\bibfield
   {title} {\bibinfo {title} {Optimal control of coupled spin dynamics: design
  of nmr pulse sequences by gradient ascent algorithms},\ }\href
  {https://doi.org/10.1016/j.jmr.2004.11.004} {\bibfield  {journal} {\bibinfo
  {journal} {J. Magn. Reson.}\ }\textbf {\bibinfo {volume} {172}},\ \bibinfo
  {pages} {296} (\bibinfo {year} {2005})}\BibitemShut {NoStop}%
\bibitem [{\citenamefont {Zahedinejad}\ \emph {et~al.}(2016)\citenamefont
  {Zahedinejad}, \citenamefont {Ghosh},\ and\ \citenamefont {Sanders}}]{ZG16}%
  \BibitemOpen
  \bibfield  {author} {\bibinfo {author} {\bibfnamefont {E.}~\bibnamefont
  {Zahedinejad}}, \bibinfo {author} {\bibfnamefont {J.}~\bibnamefont {Ghosh}},\
  and\ \bibinfo {author} {\bibfnamefont {B.~C.}\ \bibnamefont {Sanders}},\
  }\bibfield  {title} {\bibinfo {title} {Designing High-Fidelity Single-Shot
  Three-Qubit Gates: A Machine-Learning Approach},\ }\href
  {https://doi.org/10.1103/PhysRevApplied.6.054005} {\bibfield  {journal}
  {\bibinfo  {journal} {Phys. Rev. Applied}\ }\textbf {\bibinfo {volume} {6}},\
  \bibinfo {pages} {054005} (\bibinfo {year} {2016})}\BibitemShut {NoStop}%
\bibitem [{\citenamefont {Yang}\ \emph {et~al.}(2019)\citenamefont {Yang},
  \citenamefont {Li},\ and\ \citenamefont {Peng}}]{YL19}%
  \BibitemOpen
  \bibfield  {author} {\bibinfo {author} {\bibfnamefont {X.}~\bibnamefont
  {Yang}}, \bibinfo {author} {\bibfnamefont {J.}~\bibnamefont {Li}},\ and\
  \bibinfo {author} {\bibfnamefont {X.}~\bibnamefont {Peng}},\ }\bibfield
  {title} {\bibinfo {title} {An improved differential evolution algorithm for
  learning high-fidelity quantum controls},\ }\href
  {https://doi.org/https://doi.org/10.1016/j.scib.2019.07.013} {\bibfield
  {journal} {\bibinfo  {journal} {Sci. Bull.}\ }\textbf {\bibinfo {volume}
  {64}},\ \bibinfo {pages} {1402 } (\bibinfo {year} {2019})}\BibitemShut
  {NoStop}%
\bibitem [{\citenamefont {Riaz}\ \emph {et~al.}(2019)\citenamefont {Riaz},
  \citenamefont {Shuang},\ and\ \citenamefont {Qamar}}]{RBS19}%
  \BibitemOpen
  \bibfield  {author} {\bibinfo {author} {\bibfnamefont {B.}~\bibnamefont
  {Riaz}}, \bibinfo {author} {\bibfnamefont {C.}~\bibnamefont {Shuang}},\ and\
  \bibinfo {author} {\bibfnamefont {S.}~\bibnamefont {Qamar}},\ }\bibfield
  {title} {\bibinfo {title} {Optimal control methods for quantum gate
  preparation: a comparative study},\ }\href
  {https://doi.org/10.1007/s11128-019-2190-0} {\bibfield  {journal} {\bibinfo
  {journal} {Quantum Inf. Proc.}\ }\textbf {\bibinfo {volume} {18}},\ \bibinfo
  {pages} {100} (\bibinfo {year} {2019})}\BibitemShut {NoStop}%
\bibitem [{\citenamefont {Moler}\ and\ \citenamefont {Van~Loan}(2003)}]{MCV03}%
  \BibitemOpen
  \bibfield  {author} {\bibinfo {author} {\bibfnamefont {C.}~\bibnamefont
  {Moler}}\ and\ \bibinfo {author} {\bibfnamefont {C.}~\bibnamefont
  {Van~Loan}},\ }\bibfield  {title} {\bibinfo {title} {Nineteen Dubious Ways to
  Compute the Exponential of a Matrix, Twenty-Five Years Later},\ }\href
  {https://doi.org/10.1137/s00361445024180} {\bibfield  {journal} {\bibinfo
  {journal} {SIAM review}\ }\textbf {\bibinfo {volume} {45}},\ \bibinfo {pages}
  {3} (\bibinfo {year} {2003})}\BibitemShut {NoStop}%
\bibitem [{\citenamefont {Preskill}(2018)}]{PJN19}%
  \BibitemOpen
  \bibfield  {author} {\bibinfo {author} {\bibfnamefont {J.}~\bibnamefont
  {Preskill}},\ }\bibfield  {title} {\bibinfo {title} {Quantum {C}omputing in
  the {NISQ} era and beyond},\ }\href
  {https://doi.org/10.22331/q-2018-08-06-79} {\bibfield  {journal} {\bibinfo
  {journal} {{Quantum}}\ }\textbf {\bibinfo {volume} {2}},\ \bibinfo {pages}
  {79} (\bibinfo {year} {2018})}\BibitemShut {NoStop}%
\bibitem [{\citenamefont {White}(1992)}]{WSR92}%
  \BibitemOpen
  \bibfield  {author} {\bibinfo {author} {\bibfnamefont {S.~R.}\ \bibnamefont
  {White}},\ }\bibfield  {title} {\bibinfo {title} {Density matrix formulation
  for quantum renormalization groups},\ }\href
  {https://doi.org/10.1103/PhysRevLett.69.2863} {\bibfield  {journal} {\bibinfo
   {journal} {Phys. Rev. Lett.}\ }\textbf {\bibinfo {volume} {69}},\ \bibinfo
  {pages} {2863} (\bibinfo {year} {1992})}\BibitemShut {NoStop}%
\bibitem [{\citenamefont {Vidal}(2003)}]{VGE03}%
  \BibitemOpen
  \bibfield  {author} {\bibinfo {author} {\bibfnamefont {G.}~\bibnamefont
  {Vidal}},\ }\bibfield  {title} {\bibinfo {title} {{Efficient Classical
  Simulation of Slightly Entangled Quantum Computations}},\ }\href
  {https://doi.org/10.1103/PhysRevLett.91.147902} {\bibfield  {journal}
  {\bibinfo  {journal} {Phys. Rev. Lett.}\ }\textbf {\bibinfo {volume} {91}},\
  \bibinfo {pages} {147902} (\bibinfo {year} {2003})}\BibitemShut {NoStop}%
\bibitem [{\citenamefont {Evenbly}\ and\ \citenamefont {Vidal}(2015)}]{EGG15}%
  \BibitemOpen
  \bibfield  {author} {\bibinfo {author} {\bibfnamefont {G.}~\bibnamefont
  {Evenbly}}\ and\ \bibinfo {author} {\bibfnamefont {G.}~\bibnamefont
  {Vidal}},\ }\bibfield  {title} {\bibinfo {title} {{Tensor Network
  Renormalization}},\ }\href {https://doi.org/10.1103/PhysRevLett.115.180405}
  {\bibfield  {journal} {\bibinfo  {journal} {Phys. Rev. Lett.}\ }\textbf
  {\bibinfo {volume} {115}},\ \bibinfo {pages} {180405} (\bibinfo {year}
  {2015})}\BibitemShut {NoStop}%
\bibitem [{\citenamefont {Doria}\ \emph {et~al.}(2011)\citenamefont {Doria},
  \citenamefont {Calarco},\ and\ \citenamefont {Montangero}}]{DPC}%
  \BibitemOpen
  \bibfield  {author} {\bibinfo {author} {\bibfnamefont {P.}~\bibnamefont
  {Doria}}, \bibinfo {author} {\bibfnamefont {T.}~\bibnamefont {Calarco}},\
  and\ \bibinfo {author} {\bibfnamefont {S.}~\bibnamefont {Montangero}},\
  }\bibfield  {title} {\bibinfo {title} {{Optimal Control Technique for
  Many-Body Quantum Dynamics}},\ }\href
  {https://doi.org/10.1103/PhysRevLett.106.190501} {\bibfield  {journal}
  {\bibinfo  {journal} {Phys. Rev. Lett.}\ }\textbf {\bibinfo {volume} {106}},\
  \bibinfo {pages} {190501} (\bibinfo {year} {2011})}\BibitemShut {NoStop}%
\bibitem [{\citenamefont {Ryan}\ \emph {et~al.}(2008)\citenamefont {Ryan},
  \citenamefont {Negrevergne}, \citenamefont {Laforest}, \citenamefont
  {Knill},\ and\ \citenamefont {Laflamme}}]{RCN08}%
  \BibitemOpen
  \bibfield  {author} {\bibinfo {author} {\bibfnamefont {C.~A.}\ \bibnamefont
  {Ryan}}, \bibinfo {author} {\bibfnamefont {C.}~\bibnamefont {Negrevergne}},
  \bibinfo {author} {\bibfnamefont {M.}~\bibnamefont {Laforest}}, \bibinfo
  {author} {\bibfnamefont {E.}~\bibnamefont {Knill}},\ and\ \bibinfo {author}
  {\bibfnamefont {R.}~\bibnamefont {Laflamme}},\ }\bibfield  {title} {\bibinfo
  {title} {Liquid-state nuclear magnetic resonance as a testbed for developing
  quantum control methods},\ }\href
  {https://doi.org/10.1103/PhysRevA.78.012328} {\bibfield  {journal} {\bibinfo
  {journal} {Phys. Rev. A}\ }\textbf {\bibinfo {volume} {78}},\ \bibinfo
  {pages} {012328} (\bibinfo {year} {2008})}\BibitemShut {NoStop}%
\bibitem [{\citenamefont {Li}()}]{LJ19}%
  \BibitemOpen
  \bibfield  {author} {\bibinfo {author} {\bibfnamefont {J.}~\bibnamefont
  {Li}},\ }\bibfield  {title} {\bibinfo {title} {Subsystem-based approach to
  scalable quantum optimal control},\ }\href@noop {} {\ }\Eprint
  {https://arxiv.org/abs/1910.02061} {arXiv:1910.02061} \BibitemShut {NoStop}%
\bibitem [{\citenamefont {Berry}\ \emph {et~al.}(2015)\citenamefont {Berry},
  \citenamefont {Childs}, \citenamefont {Cleve}, \citenamefont {Kothari},\ and\
  \citenamefont {Somma}}]{BDW15}%
  \BibitemOpen
  \bibfield  {author} {\bibinfo {author} {\bibfnamefont {D.~W.}\ \bibnamefont
  {Berry}}, \bibinfo {author} {\bibfnamefont {A.~M.}\ \bibnamefont {Childs}},
  \bibinfo {author} {\bibfnamefont {R.}~\bibnamefont {Cleve}}, \bibinfo
  {author} {\bibfnamefont {R.}~\bibnamefont {Kothari}},\ and\ \bibinfo {author}
  {\bibfnamefont {R.~D.}\ \bibnamefont {Somma}},\ }\bibfield  {title} {\bibinfo
  {title} {Simulating Hamiltonian Dynamics with a Truncated Taylor Series},\
  }\href {https://doi.org/10.1103/PhysRevLett.114.090502} {\bibfield  {journal}
  {\bibinfo  {journal} {Phys. Rev. Lett.}\ }\textbf {\bibinfo {volume} {114}},\
  \bibinfo {pages} {090502} (\bibinfo {year} {2015})}\BibitemShut {NoStop}%
\bibitem [{\citenamefont {Kieferov\'a}\ \emph {et~al.}(2019)\citenamefont
  {Kieferov\'a}, \citenamefont {Scherer},\ and\ \citenamefont {Berry}}]{KMS19}%
  \BibitemOpen
  \bibfield  {author} {\bibinfo {author} {\bibfnamefont {M.}~\bibnamefont
  {Kieferov\'a}}, \bibinfo {author} {\bibfnamefont {A.}~\bibnamefont
  {Scherer}},\ and\ \bibinfo {author} {\bibfnamefont {D.~W.}\ \bibnamefont
  {Berry}},\ }\bibfield  {title} {\bibinfo {title} {Simulating the dynamics of
  time-dependent hamiltonians with a truncated dyson series},\ }\href
  {https://doi.org/10.1103/PhysRevA.99.042314} {\bibfield  {journal} {\bibinfo
  {journal} {Phys. Rev. A}\ }\textbf {\bibinfo {volume} {99}},\ \bibinfo
  {pages} {042314} (\bibinfo {year} {2019})}\BibitemShut {NoStop}%
\bibitem [{\citenamefont {Gradl}\ \emph {et~al.}(2006)\citenamefont {Gradl},
  \citenamefont {Sp{\"o}rl}, \citenamefont {Huckle}, \citenamefont {Glaser},\
  and\ \citenamefont {Schulte-Herbr{\"u}ggen}}]{GTS06}%
  \BibitemOpen
  \bibfield  {author} {\bibinfo {author} {\bibfnamefont {T.}~\bibnamefont
  {Gradl}}, \bibinfo {author} {\bibfnamefont {A.}~\bibnamefont {Sp{\"o}rl}},
  \bibinfo {author} {\bibfnamefont {T.}~\bibnamefont {Huckle}}, \bibinfo
  {author} {\bibfnamefont {S.~J.}\ \bibnamefont {Glaser}},\ and\ \bibinfo
  {author} {\bibfnamefont {T.}~\bibnamefont {Schulte-Herbr{\"u}ggen}},\
  }\bibfield  {title} {\bibinfo {title} {Parallelising matrix operations on
  clusters for an optimal control-based quantum compiler},\ }in\ \href
  {https://doi.org/10.1007/11823285_78} {\emph {\bibinfo {booktitle} {European
  Conference on Parallel Processing}}}\ (\bibinfo {organization} {Springer},\
  \bibinfo {year} {2006})\ pp.\ \bibinfo {pages} {751--762}\BibitemShut
  {NoStop}%
\bibitem [{\citenamefont {Edwards}\ and\ \citenamefont {Kuprov}(2012)}]{ELJ12}%
  \BibitemOpen
  \bibfield  {author} {\bibinfo {author} {\bibfnamefont {L.~J.}\ \bibnamefont
  {Edwards}}\ and\ \bibinfo {author} {\bibfnamefont {I.}~\bibnamefont
  {Kuprov}},\ }\bibfield  {title} {\bibinfo {title} {Parallel density matrix
  propagation in spin dynamics simulations},\ }\href
  {https://doi.org/10.1063/1.3679656} {\bibfield  {journal} {\bibinfo
  {journal} {J. Chem. Phys.}\ }\textbf {\bibinfo {volume} {136}},\ \bibinfo
  {pages} {044108} (\bibinfo {year} {2012})}\BibitemShut {NoStop}%
\bibitem [{\citenamefont {Riahi}\ \emph {et~al.}(2016)\citenamefont {Riahi},
  \citenamefont {Salomon}, \citenamefont {Glaser},\ and\ \citenamefont
  {Sugny}}]{RMK16}%
  \BibitemOpen
  \bibfield  {author} {\bibinfo {author} {\bibfnamefont {M.~K.}\ \bibnamefont
  {Riahi}}, \bibinfo {author} {\bibfnamefont {J.}~\bibnamefont {Salomon}},
  \bibinfo {author} {\bibfnamefont {S.~J.}\ \bibnamefont {Glaser}},\ and\
  \bibinfo {author} {\bibfnamefont {D.}~\bibnamefont {Sugny}},\ }\bibfield
  {title} {\bibinfo {title} {Fully efficient time-parallelized quantum optimal
  control algorithm},\ }\href {https://doi.org/10.1103/PhysRevA.93.043410}
  {\bibfield  {journal} {\bibinfo  {journal} {Phys. Rev. A}\ }\textbf {\bibinfo
  {volume} {93}},\ \bibinfo {pages} {043410} (\bibinfo {year}
  {2016})}\BibitemShut {NoStop}%
\bibitem [{\citenamefont {Li}\ \emph {et~al.}(2017)\citenamefont {Li},
  \citenamefont {Yang}, \citenamefont {Peng},\ and\ \citenamefont
  {Sun}}]{LYP17}%
  \BibitemOpen
  \bibfield  {author} {\bibinfo {author} {\bibfnamefont {J.}~\bibnamefont
  {Li}}, \bibinfo {author} {\bibfnamefont {X.}~\bibnamefont {Yang}}, \bibinfo
  {author} {\bibfnamefont {X.}~\bibnamefont {Peng}},\ and\ \bibinfo {author}
  {\bibfnamefont {C.-P.}\ \bibnamefont {Sun}},\ }\bibfield  {title} {\bibinfo
  {title} {{Hybrid Quantum-Classical Approach to Quantum Optimal Control}},\
  }\href {https://doi.org/10.1103/physrevlett.118.150503} {\bibfield  {journal}
  {\bibinfo  {journal} {Phys. Rev. Lett.}\ }\textbf {\bibinfo {volume} {118}},\
  \bibinfo {pages} {150503} (\bibinfo {year} {2017})}\BibitemShut {NoStop}%
\bibitem [{\citenamefont {Lu}\ \emph {et~al.}(2017)\citenamefont {Lu},
  \citenamefont {Li}, \citenamefont {Li}, \citenamefont {Katiyar},
  \citenamefont {Park}, \citenamefont {Feng}, \citenamefont {Xin},
  \citenamefont {Li}, \citenamefont {Long}, \citenamefont {Brodutch},
  \citenamefont {Zeng},\ and\ \citenamefont {Laflamme}}]{LLL17}%
  \BibitemOpen
  \bibfield  {author} {\bibinfo {author} {\bibfnamefont {D.}~\bibnamefont
  {Lu}}, \bibinfo {author} {\bibfnamefont {K.}~\bibnamefont {Li}}, \bibinfo
  {author} {\bibfnamefont {J.}~\bibnamefont {Li}}, \bibinfo {author}
  {\bibfnamefont {H.}~\bibnamefont {Katiyar}}, \bibinfo {author} {\bibfnamefont
  {A.~J.}\ \bibnamefont {Park}}, \bibinfo {author} {\bibfnamefont
  {G.}~\bibnamefont {Feng}}, \bibinfo {author} {\bibfnamefont {T.}~\bibnamefont
  {Xin}}, \bibinfo {author} {\bibfnamefont {H.}~\bibnamefont {Li}}, \bibinfo
  {author} {\bibfnamefont {G.}~\bibnamefont {Long}}, \bibinfo {author}
  {\bibfnamefont {A.}~\bibnamefont {Brodutch}}, \bibinfo {author}
  {\bibfnamefont {B.}~\bibnamefont {Zeng}},\ and\ \bibinfo {author}
  {\bibfnamefont {R.}~\bibnamefont {Laflamme}},\ }\bibfield  {title} {\bibinfo
  {title} {Enhancing quantum control by bootstrapping a quantum processor of 12
  qubits},\ }\href {https://doi.org/10.1038/s41534-017-0045-z} {\bibfield
  {journal} {\bibinfo  {journal} {npj Quantum Inf.}\ }\textbf {\bibinfo
  {volume} {3}},\ \bibinfo {pages} {45} (\bibinfo {year} {2017})}\BibitemShut
  {NoStop}%
\bibitem [{\citenamefont {Janke}\ and\ \citenamefont {Sauer}(1992)}]{JWS92}%
  \BibitemOpen
  \bibfield  {author} {\bibinfo {author} {\bibfnamefont {W.}~\bibnamefont
  {Janke}}\ and\ \bibinfo {author} {\bibfnamefont {T.}~\bibnamefont {Sauer}},\
  }\bibfield  {title} {\bibinfo {title} {Properties of higher-order trotter
  formulas},\ }\href {https://doi.org/10.1016/0375-9601(92)90035-k} {\bibfield
  {journal} {\bibinfo  {journal} {Phys. Lett. A}\ }\textbf {\bibinfo {volume}
  {165}},\ \bibinfo {pages} {199} (\bibinfo {year} {1992})}\BibitemShut
  {NoStop}%
\bibitem [{\citenamefont {Kobayashi}\ \emph {et~al.}(1994)\citenamefont
  {Kobayashi}, \citenamefont {Hatano},\ and\ \citenamefont {Suzuki}}]{KHH94}%
  \BibitemOpen
  \bibfield  {author} {\bibinfo {author} {\bibfnamefont {H.}~\bibnamefont
  {Kobayashi}}, \bibinfo {author} {\bibfnamefont {N.}~\bibnamefont {Hatano}},\
  and\ \bibinfo {author} {\bibfnamefont {M.}~\bibnamefont {Suzuki}},\
  }\bibfield  {title} {\bibinfo {title} {Study of correction terms for
  higher-order decompositions of exponential operators},\ }\href
  {https://doi.org/10.1016/0378-4371(94)00181-2} {\bibfield  {journal}
  {\bibinfo  {journal} {Physica A}\ }\textbf {\bibinfo {volume} {211}},\
  \bibinfo {pages} {234} (\bibinfo {year} {1994})}\BibitemShut {NoStop}%
\bibitem [{\citenamefont {Hatano}\ and\ \citenamefont {Suzuki}(2005)}]{HNS05}%
  \BibitemOpen
  \bibfield  {author} {\bibinfo {author} {\bibfnamefont {N.}~\bibnamefont
  {Hatano}}\ and\ \bibinfo {author} {\bibfnamefont {M.}~\bibnamefont
  {Suzuki}},\ }\bibfield  {title} {\bibinfo {title} {Finding exponential
  product formulas of higher orders},\ }in\ \href
  {https://doi.org/10.1007/11526216_2} {\emph {\bibinfo {booktitle} {Quantum
  annealing and other optimization methods}}}\ (\bibinfo  {publisher}
  {Springer},\ \bibinfo {year} {2005})\ pp.\ \bibinfo {pages}
  {37--68}\BibitemShut {NoStop}%
\bibitem [{\citenamefont {Bhole}\ and\ \citenamefont {Jones}(2018)}]{BJ18}%
  \BibitemOpen
  \bibfield  {author} {\bibinfo {author} {\bibfnamefont {G.}~\bibnamefont
  {Bhole}}\ and\ \bibinfo {author} {\bibfnamefont {J.~A.}\ \bibnamefont
  {Jones}},\ }\bibfield  {title} {\bibinfo {title} {Practical pulse
  engineering: Gradient ascent without matrix exponentiation},\ }\href
  {https://doi.org/10.1007/s11467-018-0791-1} {\bibfield  {journal} {\bibinfo
  {journal} {Front. Phys.}\ }\textbf {\bibinfo {volume} {13}},\ \bibinfo
  {pages} {130312} (\bibinfo {year} {2018})}\BibitemShut {NoStop}%
\bibitem [{\citenamefont {Suzuki}(1976)}]{suzuki1976generalized}%
  \BibitemOpen
  \bibfield  {author} {\bibinfo {author} {\bibfnamefont {M.}~\bibnamefont
  {Suzuki}},\ }\bibfield  {title} {\bibinfo {title} {Generalized trotter's
  formula and systematic approximants of exponential operators and inner
  derivations with applications to many-body problems},\ }\href
  {https://doi.org/10.1007/BF01609348} {\bibfield  {journal} {\bibinfo
  {journal} {Commun.Math. Phys.}\ }\textbf {\bibinfo {volume} {51}},\ \bibinfo
  {pages} {183} (\bibinfo {year} {1976})}\BibitemShut {NoStop}%
\bibitem [{\citenamefont {Suzuki}(1985)}]{suzuki1985decomposition}%
  \BibitemOpen
  \bibfield  {author} {\bibinfo {author} {\bibfnamefont {M.}~\bibnamefont
  {Suzuki}},\ }\bibfield  {title} {\bibinfo {title} {Decomposition formulas of
  exponential operators and lie exponentials with some applications to quantum
  mechanics and statistical physics},\ }\href
  {https://doi.org/10.1063/1.526596} {\bibfield  {journal} {\bibinfo  {journal}
  {J. Math. Phys.}\ }\textbf {\bibinfo {volume} {26}},\ \bibinfo {pages} {601}
  (\bibinfo {year} {1985})}\BibitemShut {NoStop}%
\bibitem [{\citenamefont {Georgescu}\ \emph {et~al.}(2014)\citenamefont
  {Georgescu}, \citenamefont {Ashhab},\ and\ \citenamefont
  {Nori}}]{RevModPhys.86.153}%
  \BibitemOpen
  \bibfield  {author} {\bibinfo {author} {\bibfnamefont {I.~M.}\ \bibnamefont
  {Georgescu}}, \bibinfo {author} {\bibfnamefont {S.}~\bibnamefont {Ashhab}},\
  and\ \bibinfo {author} {\bibfnamefont {F.}~\bibnamefont {Nori}},\ }\bibfield
  {title} {\bibinfo {title} {Quantum simulation},\ }\href
  {https://doi.org/10.1103/RevModPhys.86.153} {\bibfield  {journal} {\bibinfo
  {journal} {Rev. Mod. Phys.}\ }\textbf {\bibinfo {volume} {86}},\ \bibinfo
  {pages} {153} (\bibinfo {year} {2014})}\BibitemShut {NoStop}%
\bibitem [{\citenamefont {Childs}\ \emph {et~al.}(2021)\citenamefont {Childs},
  \citenamefont {Su}, \citenamefont {Tran}, \citenamefont {Wiebe},\ and\
  \citenamefont {Zhu}}]{PhysRevX.11.011020}%
  \BibitemOpen
  \bibfield  {author} {\bibinfo {author} {\bibfnamefont {A.~M.}\ \bibnamefont
  {Childs}}, \bibinfo {author} {\bibfnamefont {Y.}~\bibnamefont {Su}}, \bibinfo
  {author} {\bibfnamefont {M.~C.}\ \bibnamefont {Tran}}, \bibinfo {author}
  {\bibfnamefont {N.}~\bibnamefont {Wiebe}},\ and\ \bibinfo {author}
  {\bibfnamefont {S.}~\bibnamefont {Zhu}},\ }\bibfield  {title} {\bibinfo
  {title} {Theory of Trotter Error with Commutator Scaling},\ }\href
  {https://doi.org/10.1103/PhysRevX.11.011020} {\bibfield  {journal} {\bibinfo
  {journal} {Phys. Rev. X}\ }\textbf {\bibinfo {volume} {11}},\ \bibinfo
  {pages} {011020} (\bibinfo {year} {2021})}\BibitemShut {NoStop}%
\bibitem [{\citenamefont {Suzuki}(1992{\natexlab{a}})}]{suzuki1992sym}%
  \BibitemOpen
  \bibfield  {author} {\bibinfo {author} {\bibfnamefont {M.}~\bibnamefont
  {Suzuki}},\ }\bibfield  {title} {\bibinfo {title} {General theory of
  higher-order decomposition of exponential operators and symplectic
  integrators},\ }\href {https://doi.org/10.1016/0375-9601(92)90335-J}
  {\bibfield  {journal} {\bibinfo  {journal} {Phys. Lett. A}\ }\textbf
  {\bibinfo {volume} {165}},\ \bibinfo {pages} {387} (\bibinfo {year}
  {1992}{\natexlab{a}})}\BibitemShut {NoStop}%
\bibitem [{\citenamefont {Ruth}(1983)}]{ruth1983canonical}%
  \BibitemOpen
  \bibfield  {author} {\bibinfo {author} {\bibfnamefont {R.~D.}\ \bibnamefont
  {Ruth}},\ }\bibfield  {title} {\bibinfo {title} {A canonical integration
  technique},\ }\href {https://doi.org/10.1109/TNS.1983.4332919} {\bibfield
  {journal} {\bibinfo  {journal} {IEEE Trans. Nucl. Sci.}\ }\textbf {\bibinfo
  {volume} {30}},\ \bibinfo {pages} {2669} (\bibinfo {year}
  {1983})}\BibitemShut {NoStop}%
\bibitem [{\citenamefont {Suzuki}(1992{\natexlab{b}})}]{suzuki1992general}%
  \BibitemOpen
  \bibfield  {author} {\bibinfo {author} {\bibfnamefont {M.}~\bibnamefont
  {Suzuki}},\ }\bibfield  {title} {\bibinfo {title} {General nonsymmetric
  higher-order decomposition of exponential operators and symplectic
  integrators},\ }\href {https://doi.org/10.1143/JPSJ.61.3015} {\bibfield
  {journal} {\bibinfo  {journal} {J. Phys. Soc. Japan}\ }\textbf {\bibinfo
  {volume} {61}},\ \bibinfo {pages} {3015} (\bibinfo {year}
  {1992}{\natexlab{b}})}\BibitemShut {NoStop}%
\bibitem [{\citenamefont {Liu}\ \emph {et~al.}(2020)\citenamefont {Liu},
  \citenamefont {Hines}, \citenamefont {Li}, \citenamefont {Ajoy},\ and\
  \citenamefont {Cappellaro}}]{PhysRevA.102.010601}%
  \BibitemOpen
  \bibfield  {author} {\bibinfo {author} {\bibfnamefont {Y.-X.}\ \bibnamefont
  {Liu}}, \bibinfo {author} {\bibfnamefont {J.}~\bibnamefont {Hines}}, \bibinfo
  {author} {\bibfnamefont {Z.}~\bibnamefont {Li}}, \bibinfo {author}
  {\bibfnamefont {A.}~\bibnamefont {Ajoy}},\ and\ \bibinfo {author}
  {\bibfnamefont {P.}~\bibnamefont {Cappellaro}},\ }\bibfield  {title}
  {\bibinfo {title} {High-fidelity trotter formulas for digital quantum
  simulation},\ }\href {https://doi.org/10.1103/PhysRevA.102.010601} {\bibfield
   {journal} {\bibinfo  {journal} {Phys. Rev. A}\ }\textbf {\bibinfo {volume}
  {102}},\ \bibinfo {pages} {010601} (\bibinfo {year} {2020})}\BibitemShut
  {NoStop}%
\bibitem [{\citenamefont {Stothers}(2010)}]{stothers2010complexity}%
  \BibitemOpen
  \bibfield  {author} {\bibinfo {author} {\bibfnamefont {A.~J.}\ \bibnamefont
  {Stothers}},\ }\bibfield  {title} {\bibinfo {title} {On the complexity of
  matrix multiplication (Ph.D. Thesis, the university of Edinburgh, 2010)}
  } \BibitemShut {NoStop}%
\bibitem [{\citenamefont {Friedenauer}\ \emph {et~al.}(2008)\citenamefont
  {Friedenauer}, \citenamefont {Schmitz}, \citenamefont {Glueckert},
  \citenamefont {Porras},\ and\ \citenamefont {Sch{\"a}tz}}]{FAZ08}%
  \BibitemOpen
  \bibfield  {author} {\bibinfo {author} {\bibfnamefont {A.}~\bibnamefont
  {Friedenauer}}, \bibinfo {author} {\bibfnamefont {H.}~\bibnamefont
  {Schmitz}}, \bibinfo {author} {\bibfnamefont {J.~T.}\ \bibnamefont
  {Glueckert}}, \bibinfo {author} {\bibfnamefont {D.}~\bibnamefont {Porras}},\
  and\ \bibinfo {author} {\bibfnamefont {T.}~\bibnamefont {Sch{\"a}tz}},\
  }\bibfield  {title} {\bibinfo {title} {Simulating a quantum magnet with
  trapped ions},\ }\href {https://doi.org/10.1038/nature09071} {\bibfield
  {journal} {\bibinfo  {journal} {Nat. Phys.}\ }\textbf {\bibinfo {volume}
  {4}},\ \bibinfo {pages} {757} (\bibinfo {year} {2008})}\BibitemShut {NoStop}%
\bibitem [{\citenamefont {Zhang}\ \emph {et~al.}(2011)\citenamefont {Zhang},
  \citenamefont {Ryan}, \citenamefont {Laflamme},\ and\ \citenamefont
  {Baugh}}]{ZYR11}%
  \BibitemOpen
  \bibfield  {author} {\bibinfo {author} {\bibfnamefont {Y.}~\bibnamefont
  {Zhang}}, \bibinfo {author} {\bibfnamefont {C.~A.}\ \bibnamefont {Ryan}},
  \bibinfo {author} {\bibfnamefont {R.}~\bibnamefont {Laflamme}},\ and\
  \bibinfo {author} {\bibfnamefont {J.}~\bibnamefont {Baugh}},\ }\bibfield
  {title} {\bibinfo {title} {Coherent Control of Two Nuclear Spins Using the
  Anisotropic Hyperfine Interaction},\ }\href
  {https://doi.org/10.1103/PhysRevLett.107.170503} {\bibfield  {journal}
  {\bibinfo  {journal} {Phys. Rev. Lett.}\ }\textbf {\bibinfo {volume} {107}},\
  \bibinfo {pages} {170503} (\bibinfo {year} {2011})}\BibitemShut {NoStop}%
\bibitem [{\citenamefont {Waldherr}\ \emph {et~al.}(2014)\citenamefont
  {Waldherr}, \citenamefont {Wang}, \citenamefont {Zaiser}, \citenamefont
  {Jamali}, \citenamefont {Schulte-Herbr{\"u}ggen}, \citenamefont {Abe},
  \citenamefont {Ohshima}, \citenamefont {Isoya}, \citenamefont {Du},
  \citenamefont {Neumann},\ and\ \citenamefont {Wrachtrup}}]{WGW14}%
  \BibitemOpen
  \bibfield  {author} {\bibinfo {author} {\bibfnamefont {G.}~\bibnamefont
  {Waldherr}}, \bibinfo {author} {\bibfnamefont {Y.}~\bibnamefont {Wang}},
  \bibinfo {author} {\bibfnamefont {S.}~\bibnamefont {Zaiser}}, \bibinfo
  {author} {\bibfnamefont {M.}~\bibnamefont {Jamali}}, \bibinfo {author}
  {\bibfnamefont {T.}~\bibnamefont {Schulte-Herbr{\"u}ggen}}, \bibinfo {author}
  {\bibfnamefont {H.}~\bibnamefont {Abe}}, \bibinfo {author} {\bibfnamefont
  {T.}~\bibnamefont {Ohshima}}, \bibinfo {author} {\bibfnamefont
  {J.}~\bibnamefont {Isoya}}, \bibinfo {author} {\bibfnamefont
  {J.}~\bibnamefont {Du}}, \bibinfo {author} {\bibfnamefont {P.}~\bibnamefont
  {Neumann}},\ and\ \bibinfo {author} {\bibfnamefont {J.}~\bibnamefont
  {Wrachtrup}},\ }\bibfield  {title} {\bibinfo {title} {Quantum error
  correction in a solid-state hybrid spin register},\ }\href
  {https://doi.org/10.1038/nature12919} {\bibfield  {journal} {\bibinfo
  {journal} {Nature}\ }\textbf {\bibinfo {volume} {506}},\ \bibinfo {pages}
  {204} (\bibinfo {year} {2014})}\BibitemShut {NoStop}%
\bibitem [{\citenamefont {Dolde}\ \emph {et~al.}(2014)\citenamefont {Dolde},
  \citenamefont {Bergholm}, \citenamefont {Wang}, \citenamefont {Jakobi},
  \citenamefont {Naydenov}, \citenamefont {Pezzagna}, \citenamefont {Meijer},
  \citenamefont {Jelezko}, \citenamefont {Neumann}, \citenamefont
  {Schulte-Herbr{\"u}ggen}, \citenamefont {Biamonte},\ and\ \citenamefont
  {Wrachtrup}}]{DFB14}%
  \BibitemOpen
  \bibfield  {author} {\bibinfo {author} {\bibfnamefont {F.}~\bibnamefont
  {Dolde}}, \bibinfo {author} {\bibfnamefont {V.}~\bibnamefont {Bergholm}},
  \bibinfo {author} {\bibfnamefont {Y.}~\bibnamefont {Wang}}, \bibinfo {author}
  {\bibfnamefont {I.}~\bibnamefont {Jakobi}}, \bibinfo {author} {\bibfnamefont
  {B.}~\bibnamefont {Naydenov}}, \bibinfo {author} {\bibfnamefont
  {S.}~\bibnamefont {Pezzagna}}, \bibinfo {author} {\bibfnamefont
  {J.}~\bibnamefont {Meijer}}, \bibinfo {author} {\bibfnamefont
  {F.}~\bibnamefont {Jelezko}}, \bibinfo {author} {\bibfnamefont
  {P.}~\bibnamefont {Neumann}}, \bibinfo {author} {\bibfnamefont
  {T.}~\bibnamefont {Schulte-Herbr{\"u}ggen}}, \bibinfo {author} {\bibfnamefont
  {J.}~\bibnamefont {Biamonte}},\ and\ \bibinfo {author} {\bibfnamefont
  {J.}~\bibnamefont {Wrachtrup}},\ }\bibfield  {title} {\bibinfo {title}
  {High-fidelity spin entanglement using optimal control},\ }\href
  {https://doi.org/10.1038/ncomms4371} {\bibfield  {journal} {\bibinfo
  {journal} {Nat. Commun.}\ }\textbf {\bibinfo {volume} {5}},\ \bibinfo {pages}
  {3371} (\bibinfo {year} {2014})}\BibitemShut {NoStop}%
\bibitem [{\citenamefont {Motzoi}\ \emph {et~al.}(2009)\citenamefont {Motzoi},
  \citenamefont {Gambetta}, \citenamefont {Rebentrost},\ and\ \citenamefont
  {Wilhelm}}]{MFG09}%
  \BibitemOpen
  \bibfield  {author} {\bibinfo {author} {\bibfnamefont {F.}~\bibnamefont
  {Motzoi}}, \bibinfo {author} {\bibfnamefont {J.~M.}\ \bibnamefont
  {Gambetta}}, \bibinfo {author} {\bibfnamefont {P.}~\bibnamefont
  {Rebentrost}},\ and\ \bibinfo {author} {\bibfnamefont {F.~K.}\ \bibnamefont
  {Wilhelm}},\ }\bibfield  {title} {\bibinfo {title} {Simple Pulses for
  Elimination of Leakage in Weakly Nonlinear Qubits},\ }\href
  {https://doi.org/10.1103/PhysRevLett.103.110501} {\bibfield  {journal}
  {\bibinfo  {journal} {Phys. Rev. Lett.}\ }\textbf {\bibinfo {volume} {103}},\
  \bibinfo {pages} {110501} (\bibinfo {year} {2009})}\BibitemShut {NoStop}%
\bibitem [{\citenamefont {Egger}\ and\ \citenamefont {Wilhelm}(2014)}]{EDJ13}%
  \BibitemOpen
  \bibfield  {author} {\bibinfo {author} {\bibfnamefont {D.~J.}\ \bibnamefont
  {Egger}}\ and\ \bibinfo {author} {\bibfnamefont {F.~K.}\ \bibnamefont
  {Wilhelm}},\ }\bibfield  {title} {\bibinfo {title} {Optimized controlled-z
  gates for two superconducting qubits coupled through a resonator},\ }\href
  {https://doi.org/10.1088/0953-2048/27/1/014001} {\bibfield  {journal}
  {\bibinfo  {journal} {Supercond. Sci. Technol.}\ }\textbf {\bibinfo {volume}
  {27}},\ \bibinfo {pages} {014001} (\bibinfo {year} {2014})}\BibitemShut
  {NoStop}%
\bibitem [{\citenamefont {Nebendahl}\ \emph {et~al.}(2009)\citenamefont
  {Nebendahl}, \citenamefont {H\"affner},\ and\ \citenamefont {Roos}}]{NVH09}%
  \BibitemOpen
  \bibfield  {author} {\bibinfo {author} {\bibfnamefont {V.}~\bibnamefont
  {Nebendahl}}, \bibinfo {author} {\bibfnamefont {H.}~\bibnamefont
  {H\"affner}},\ and\ \bibinfo {author} {\bibfnamefont {C.~F.}\ \bibnamefont
  {Roos}},\ }\bibfield  {title} {\bibinfo {title} {Optimal control of
  entangling operations for trapped-ion quantum computing},\ }\href
  {https://doi.org/10.1103/PhysRevA.79.012312} {\bibfield  {journal} {\bibinfo
  {journal} {Phys. Rev. A}\ }\textbf {\bibinfo {volume} {79}},\ \bibinfo
  {pages} {012312} (\bibinfo {year} {2009})}\BibitemShut {NoStop}%
\bibitem [{\citenamefont {Schindler}\ \emph {et~al.}(2011)\citenamefont
  {Schindler}, \citenamefont {Barreiro}, \citenamefont {Monz}, \citenamefont
  {Nebendahl}, \citenamefont {Nigg}, \citenamefont {Chwalla}, \citenamefont
  {Hennrich},\ and\ \citenamefont {Blatt}}]{SPB11}%
  \BibitemOpen
  \bibfield  {author} {\bibinfo {author} {\bibfnamefont {P.}~\bibnamefont
  {Schindler}}, \bibinfo {author} {\bibfnamefont {J.~T.}\ \bibnamefont
  {Barreiro}}, \bibinfo {author} {\bibfnamefont {T.}~\bibnamefont {Monz}},
  \bibinfo {author} {\bibfnamefont {V.}~\bibnamefont {Nebendahl}}, \bibinfo
  {author} {\bibfnamefont {D.}~\bibnamefont {Nigg}}, \bibinfo {author}
  {\bibfnamefont {M.}~\bibnamefont {Chwalla}}, \bibinfo {author} {\bibfnamefont
  {M.}~\bibnamefont {Hennrich}},\ and\ \bibinfo {author} {\bibfnamefont
  {R.}~\bibnamefont {Blatt}},\ }\bibfield  {title} {\bibinfo {title}
  {Experimental repetitive quantum error correction},\ }\href
  {https://doi.org/10.1126/science.1203329} {\bibfield  {journal} {\bibinfo
  {journal} {Science}\ }\textbf {\bibinfo {volume} {332}},\ \bibinfo {pages}
  {1059} (\bibinfo {year} {2011})}\BibitemShut {NoStop}%
\bibitem [{\citenamefont {Saywell}\ \emph {et~al.}(2018)\citenamefont
  {Saywell}, \citenamefont {Kuprov}, \citenamefont {Goodwin}, \citenamefont
  {Carey},\ and\ \citenamefont {Freegarde}}]{SKG18}%
  \BibitemOpen
  \bibfield  {author} {\bibinfo {author} {\bibfnamefont {J.~C.}\ \bibnamefont
  {Saywell}}, \bibinfo {author} {\bibfnamefont {I.}~\bibnamefont {Kuprov}},
  \bibinfo {author} {\bibfnamefont {D.}~\bibnamefont {Goodwin}}, \bibinfo
  {author} {\bibfnamefont {M.}~\bibnamefont {Carey}},\ and\ \bibinfo {author}
  {\bibfnamefont {T.}~\bibnamefont {Freegarde}},\ }\bibfield  {title} {\bibinfo
  {title} {Optimal control of mirror pulses for cold-atom interferometry},\
  }\href {https://doi.org/10.1103/PhysRevA.98.023625} {\bibfield  {journal}
  {\bibinfo  {journal} {Phys. Rev. A}\ }\textbf {\bibinfo {volume} {98}},\
  \bibinfo {pages} {023625} (\bibinfo {year} {2018})}\BibitemShut {NoStop}%
\bibitem [{\citenamefont {Machnes}\ \emph {et~al.}(2011)\citenamefont
  {Machnes}, \citenamefont {Sander}, \citenamefont {Glaser}, \citenamefont
  {de~Fouqui\`eres}, \citenamefont {Gruslys}, \citenamefont {Schirmer},\ and\
  \citenamefont {Schulte-Herbr\"uggen}}]{PhysRevA.84.022305}%
  \BibitemOpen
  \bibfield  {author} {\bibinfo {author} {\bibfnamefont {S.}~\bibnamefont
  {Machnes}}, \bibinfo {author} {\bibfnamefont {U.}~\bibnamefont {Sander}},
  \bibinfo {author} {\bibfnamefont {S.~J.}\ \bibnamefont {Glaser}}, \bibinfo
  {author} {\bibfnamefont {P.}~\bibnamefont {de~Fouqui\`eres}}, \bibinfo
  {author} {\bibfnamefont {A.}~\bibnamefont {Gruslys}}, \bibinfo {author}
  {\bibfnamefont {S.}~\bibnamefont {Schirmer}},\ and\ \bibinfo {author}
  {\bibfnamefont {T.}~\bibnamefont {Schulte-Herbr\"uggen}},\ }\bibfield
  {title} {\bibinfo {title} {Comparing, optimizing, and benchmarking
  quantum-control algorithms in a unifying programming framework},\ }\href
  {https://doi.org/10.1103/PhysRevA.84.022305} {\bibfield  {journal} {\bibinfo
  {journal} {Phys. Rev. A}\ }\textbf {\bibinfo {volume} {84}},\ \bibinfo
  {pages} {022305} (\bibinfo {year} {2011})}\BibitemShut {NoStop}%
\bibitem [{\citenamefont {de~Fouquieres}\ \emph {et~al.}(2011)\citenamefont
  {de~Fouquieres}, \citenamefont {Schirmer}, \citenamefont {Glaser},\ and\
  \citenamefont {Kuprov}}]{de2011second}%
  \BibitemOpen
  \bibfield  {author} {\bibinfo {author} {\bibfnamefont {P.}~\bibnamefont
  {de~Fouquieres}}, \bibinfo {author} {\bibfnamefont {S.~G.}\ \bibnamefont
  {Schirmer}}, \bibinfo {author} {\bibfnamefont {S.~J.}\ \bibnamefont
  {Glaser}},\ and\ \bibinfo {author} {\bibfnamefont {I.}~\bibnamefont
  {Kuprov}},\ }\bibfield  {title} {\bibinfo {title} {Second order gradient
  ascent pulse engineering},\ }\href
  {https://doi.org/10.1016/j.jmr.2011.07.023} {\bibfield  {journal} {\bibinfo
  {journal} {J. Magn. Reson.}\ }\textbf {\bibinfo {volume} {212}},\ \bibinfo
  {pages} {412} (\bibinfo {year} {2011})}\BibitemShut {NoStop}%
\bibitem [{\citenamefont {Liu}\ and\ \citenamefont
  {Nocedal}(1989)}]{liu1989limited}%
  \BibitemOpen
  \bibfield  {author} {\bibinfo {author} {\bibfnamefont {D.~C.}\ \bibnamefont
  {Liu}}\ and\ \bibinfo {author} {\bibfnamefont {J.}~\bibnamefont {Nocedal}},\
  }\bibfield  {title} {\bibinfo {title} {On the limited memory bfgs method for
  large scale optimization},\ }\href {https://doi.org/10.1007/BF01589116}
  {\bibfield  {journal} {\bibinfo  {journal} {Math. Program.}\ }\textbf
  {\bibinfo {volume} {45}},\ \bibinfo {pages} {503} (\bibinfo {year}
  {1989})}\BibitemShut {NoStop}%
\bibitem [{\citenamefont {Pezz\`e}\ \emph {et~al.}(2018)\citenamefont
  {Pezz\`e}, \citenamefont {Smerzi}, \citenamefont {Oberthaler}, \citenamefont
  {Schmied},\ and\ \citenamefont {Treutlein}}]{PLS18}%
  \BibitemOpen
  \bibfield  {author} {\bibinfo {author} {\bibfnamefont {L.}~\bibnamefont
  {Pezz\`e}}, \bibinfo {author} {\bibfnamefont {A.}~\bibnamefont {Smerzi}},
  \bibinfo {author} {\bibfnamefont {M.~K.}\ \bibnamefont {Oberthaler}},
  \bibinfo {author} {\bibfnamefont {R.}~\bibnamefont {Schmied}},\ and\ \bibinfo
  {author} {\bibfnamefont {P.}~\bibnamefont {Treutlein}},\ }\bibfield  {title}
  {\bibinfo {title} {Quantum metrology with nonclassical states of atomic
  ensembles},\ }\href {https://doi.org/10.1103/RevModPhys.90.035005} {\bibfield
   {journal} {\bibinfo  {journal} {Rev. Mod. Phys.}\ }\textbf {\bibinfo
  {volume} {90}},\ \bibinfo {pages} {035005} (\bibinfo {year}
  {2018})}\BibitemShut {NoStop}%
\bibitem [{\citenamefont {Zhou}\ \emph {et~al.}(2020)\citenamefont {Zhou},
  \citenamefont {Ji}, \citenamefont {Nie}, \citenamefont {Yang}, \citenamefont
  {Chen}, \citenamefont {Bian},\ and\ \citenamefont {Peng}}]{ZHJ20}%
  \BibitemOpen
  \bibfield  {author} {\bibinfo {author} {\bibfnamefont {H.}~\bibnamefont
  {Zhou}}, \bibinfo {author} {\bibfnamefont {Y.}~\bibnamefont {Ji}}, \bibinfo
  {author} {\bibfnamefont {X.}~\bibnamefont {Nie}}, \bibinfo {author}
  {\bibfnamefont {X.}~\bibnamefont {Yang}}, \bibinfo {author} {\bibfnamefont
  {X.}~\bibnamefont {Chen}}, \bibinfo {author} {\bibfnamefont {J.}~\bibnamefont
  {Bian}},\ and\ \bibinfo {author} {\bibfnamefont {X.}~\bibnamefont {Peng}},\
  }\bibfield  {title} {\bibinfo {title} {Experimental Realization of Shortcuts
  to Adiabaticity in a Nonintegrable Spin Chain by Local Counterdiabatic
  Driving},\ }\href {https://doi.org/10.1103/PhysRevApplied.13.044059}
  {\bibfield  {journal} {\bibinfo  {journal} {Phys. Rev. Applied}\ }\textbf
  {\bibinfo {volume} {13}},\ \bibinfo {pages} {044059} (\bibinfo {year}
  {2020})}\BibitemShut {NoStop}%
\bibitem [{\citenamefont {Campbell}\ \emph {et~al.}(2017)\citenamefont
  {Campbell}, \citenamefont {Terhal},\ and\ \citenamefont {Vuillot}}]{CET17}%
  \BibitemOpen
  \bibfield  {author} {\bibinfo {author} {\bibfnamefont {E.~T.}\ \bibnamefont
  {Campbell}}, \bibinfo {author} {\bibfnamefont {B.~M.}\ \bibnamefont
  {Terhal}},\ and\ \bibinfo {author} {\bibfnamefont {C.}~\bibnamefont
  {Vuillot}},\ }\bibfield  {title} {\bibinfo {title} {Roads towards
  fault-tolerant universal quantum computation},\ }\href
  {https://doi.org/10.1038/nature23460} {\bibfield  {journal} {\bibinfo
  {journal} {Nature}\ }\textbf {\bibinfo {volume} {549}},\ \bibinfo {pages}
  {172} (\bibinfo {year} {2017})}\BibitemShut {NoStop}%
\bibitem [{\citenamefont {Kandala}\ \emph {et~al.}(2017)\citenamefont
  {Kandala}, \citenamefont {Mezzacapo}, \citenamefont {Temme}, \citenamefont
  {Takita}, \citenamefont {Brink}, \citenamefont {Chow},\ and\ \citenamefont
  {Gambetta}}]{KAM17}%
  \BibitemOpen
  \bibfield  {author} {\bibinfo {author} {\bibfnamefont {A.}~\bibnamefont
  {Kandala}}, \bibinfo {author} {\bibfnamefont {A.}~\bibnamefont {Mezzacapo}},
  \bibinfo {author} {\bibfnamefont {K.}~\bibnamefont {Temme}}, \bibinfo
  {author} {\bibfnamefont {M.}~\bibnamefont {Takita}}, \bibinfo {author}
  {\bibfnamefont {M.}~\bibnamefont {Brink}}, \bibinfo {author} {\bibfnamefont
  {J.~M.}\ \bibnamefont {Chow}},\ and\ \bibinfo {author} {\bibfnamefont
  {J.~M.}\ \bibnamefont {Gambetta}},\ }\bibfield  {title} {\bibinfo {title}
  {Hardware-efficient variational quantum eigensolver for small molecules and
  quantum magnets},\ }\href {https://doi.org/10.1038/nature23879} {\bibfield
  {journal} {\bibinfo  {journal} {Nature}\ }\textbf {\bibinfo {volume} {549}},\
  \bibinfo {pages} {242} (\bibinfo {year} {2017})}\BibitemShut {NoStop}%
\bibitem [{\citenamefont {Moll}\ \emph {et~al.}(2018)\citenamefont {Moll},
  \citenamefont {Barkoutsos}, \citenamefont {Bishop}, \citenamefont {Chow},
  \citenamefont {Cross}, \citenamefont {Egger}, \citenamefont {Filipp},
  \citenamefont {Fuhrer}, \citenamefont {Gambetta}, \citenamefont {Ganzhorn},
  \citenamefont {Kandala}, \citenamefont {Mezzacapo}, \citenamefont
  {M{\"u}ller}, \citenamefont {Riess}, \citenamefont {Salis}, \citenamefont
  {Smolin}, \citenamefont {Tavernelli},\ and\ \citenamefont {Temme}}]{MNB18}%
  \BibitemOpen
  \bibfield  {author} {\bibinfo {author} {\bibfnamefont {N.}~\bibnamefont
  {Moll}}, \bibinfo {author} {\bibfnamefont {P.}~\bibnamefont {Barkoutsos}},
  \bibinfo {author} {\bibfnamefont {L.~S.}\ \bibnamefont {Bishop}}, \bibinfo
  {author} {\bibfnamefont {J.~M.}\ \bibnamefont {Chow}}, \bibinfo {author}
  {\bibfnamefont {A.}~\bibnamefont {Cross}}, \bibinfo {author} {\bibfnamefont
  {D.~J.}\ \bibnamefont {Egger}}, \bibinfo {author} {\bibfnamefont
  {S.}~\bibnamefont {Filipp}}, \bibinfo {author} {\bibfnamefont
  {A.}~\bibnamefont {Fuhrer}}, \bibinfo {author} {\bibfnamefont {J.~M.}\
  \bibnamefont {Gambetta}}, \bibinfo {author} {\bibfnamefont {M.}~\bibnamefont
  {Ganzhorn}}, \bibinfo {author} {\bibfnamefont {A.}~\bibnamefont {Kandala}},
  \bibinfo {author} {\bibfnamefont {A.}~\bibnamefont {Mezzacapo}}, \bibinfo
  {author} {\bibfnamefont {P.}~\bibnamefont {M{\"u}ller}}, \bibinfo {author}
  {\bibfnamefont {W.}~\bibnamefont {Riess}}, \bibinfo {author} {\bibfnamefont
  {G.}~\bibnamefont {Salis}}, \bibinfo {author} {\bibfnamefont
  {J.}~\bibnamefont {Smolin}}, \bibinfo {author} {\bibfnamefont
  {I.}~\bibnamefont {Tavernelli}},\ and\ \bibinfo {author} {\bibfnamefont
  {K.}~\bibnamefont {Temme}},\ }\bibfield  {title} {\bibinfo {title} {Quantum
  optimization using variational algorithms on near-term quantum devices},\
  }\href {https://doi.org/10.1088/2058-9565/aab822} {\bibfield  {journal}
  {\bibinfo  {journal} {Quantum Sci. Technol.}\ }\textbf {\bibinfo {volume}
  {3}},\ \bibinfo {pages} {030503} (\bibinfo {year} {2018})}\BibitemShut
  {NoStop}%
\bibitem [{\citenamefont {Cerezo}\ \emph {et~al.}(2021)\citenamefont {Cerezo},
  \citenamefont {Arrasmith}, \citenamefont {Babbush}, \citenamefont {Benjamin},
  \citenamefont {Endo}, \citenamefont {Fujii}, \citenamefont {McClean},
  \citenamefont {Mitarai}, \citenamefont {Yuan}, \citenamefont {Cincio} \emph
  {et~al.}}]{CMA20}%
  \BibitemOpen
  \bibfield  {author} {\bibinfo {author} {\bibfnamefont {M.}~\bibnamefont
  {Cerezo}}, \bibinfo {author} {\bibfnamefont {A.}~\bibnamefont {Arrasmith}},
  \bibinfo {author} {\bibfnamefont {R.}~\bibnamefont {Babbush}}, \bibinfo
  {author} {\bibfnamefont {S.~C.}\ \bibnamefont {Benjamin}}, \bibinfo {author}
  {\bibfnamefont {S.}~\bibnamefont {Endo}}, \bibinfo {author} {\bibfnamefont
  {K.}~\bibnamefont {Fujii}}, \bibinfo {author} {\bibfnamefont {J.~R.}\
  \bibnamefont {McClean}}, \bibinfo {author} {\bibfnamefont {K.}~\bibnamefont
  {Mitarai}}, \bibinfo {author} {\bibfnamefont {X.}~\bibnamefont {Yuan}},
  \bibinfo {author} {\bibfnamefont {L.}~\bibnamefont {Cincio}}, \emph
  {et~al.},\ }\bibfield  {title} {\bibinfo {title} {Variational quantum
  algorithms},\ }\href {https://doi.org/10.1038/s42254-021-00348-9} {\bibfield
  {journal} {\bibinfo  {journal} {Nat. Rev. Phys.}\ }\textbf {\bibinfo {volume}
  {3}},\ \bibinfo {pages} {625} (\bibinfo {year} {2021})}\BibitemShut {NoStop}%  
\bibitem [{\citenamefont {Wecker}\ \emph {et~al.}(2015)\citenamefont {Wecker},
  \citenamefont {Hastings},\ and\ \citenamefont {Troyer}}]{WDH15}%
  \BibitemOpen
  \bibfield  {author} {\bibinfo {author} {\bibfnamefont {D.}~\bibnamefont
  {Wecker}}, \bibinfo {author} {\bibfnamefont {M.~B.}\ \bibnamefont
  {Hastings}},\ and\ \bibinfo {author} {\bibfnamefont {M.}~\bibnamefont
  {Troyer}},\ }\bibfield  {title} {\bibinfo {title} {Progress towards practical
  quantum variational algorithms},\ }\href
  {https://doi.org/10.1103/PhysRevA.92.042303} {\bibfield  {journal} {\bibinfo
  {journal} {Phys. Rev. A}\ }\textbf {\bibinfo {volume} {92}},\ \bibinfo
  {pages} {042303} (\bibinfo {year} {2015})}\BibitemShut {NoStop}%
\bibitem [{\citenamefont {Choquette}\ \emph {et~al.}(2021)\citenamefont
  {Choquette}, \citenamefont {Di~Paolo}, \citenamefont {Barkoutsos},
  \citenamefont {S\'en\'echal}, \citenamefont {Tavernelli},\ and\ \citenamefont
  {Blais}}]{CAD20}%
  \BibitemOpen
  \bibfield  {author} {\bibinfo {author} {\bibfnamefont {A.}~\bibnamefont
  {Choquette}}, \bibinfo {author} {\bibfnamefont {A.}~\bibnamefont {Di~Paolo}},
  \bibinfo {author} {\bibfnamefont {P.~K.}\ \bibnamefont {Barkoutsos}},
  \bibinfo {author} {\bibfnamefont {D.}~\bibnamefont {S\'en\'echal}}, \bibinfo
  {author} {\bibfnamefont {I.}~\bibnamefont {Tavernelli}},\ and\ \bibinfo
  {author} {\bibfnamefont {A.}~\bibnamefont {Blais}},\ }\bibfield  {title}
  {\bibinfo {title} {Quantum-optimal-control-inspired ansatz for variational
  quantum algorithms},\ }\href
  {https://doi.org/10.1103/PhysRevResearch.3.023092} {\bibfield  {journal}
  {\bibinfo  {journal} {Phys. Rev. Research}\ }\textbf {\bibinfo {volume}
  {3}},\ \bibinfo {pages} {023092} (\bibinfo {year} {2021})}\BibitemShut
  {NoStop}%
\bibitem [{\citenamefont {Magann}\ \emph {et~al.}(2021)\citenamefont {Magann},
  \citenamefont {Arenz}, \citenamefont {Grace}, \citenamefont {Ho},
  \citenamefont {Kosut}, \citenamefont {McClean}, \citenamefont {Rabitz},\ and\
  \citenamefont {Sarovar}}]{MAB21}%
  \BibitemOpen
  \bibfield  {author} {\bibinfo {author} {\bibfnamefont {A.~B.}\ \bibnamefont
  {Magann}}, \bibinfo {author} {\bibfnamefont {C.}~\bibnamefont {Arenz}},
  \bibinfo {author} {\bibfnamefont {M.~D.}\ \bibnamefont {Grace}}, \bibinfo
  {author} {\bibfnamefont {T.-S.}\ \bibnamefont {Ho}}, \bibinfo {author}
  {\bibfnamefont {R.~L.}\ \bibnamefont {Kosut}}, \bibinfo {author}
  {\bibfnamefont {J.~R.}\ \bibnamefont {McClean}}, \bibinfo {author}
  {\bibfnamefont {H.~A.}\ \bibnamefont {Rabitz}},\ and\ \bibinfo {author}
  {\bibfnamefont {M.}~\bibnamefont {Sarovar}},\ }\bibfield  {title} {\bibinfo
  {title} {From pulses to circuits and back again: A quantum optimal control
  perspective on variational quantum algorithms},\ }\href
  {https://doi.org/10.1103/PRXQuantum.2.010101} {\bibfield  {journal} {\bibinfo
   {journal} {PRX Quantum}\ }\textbf {\bibinfo {volume} {2}},\ \bibinfo {pages}
  {010101} (\bibinfo {year} {2021})}\BibitemShut {NoStop}%
\bibitem [{\citenamefont {Ba\~nuls}\ \emph {et~al.}(2011)\citenamefont
  {Ba\~nuls}, \citenamefont {Cirac},\ and\ \citenamefont {Hastings}}]{BMC11}%
  \BibitemOpen
  \bibfield  {author} {\bibinfo {author} {\bibfnamefont {M.~C.}\ \bibnamefont
  {Ba\~nuls}}, \bibinfo {author} {\bibfnamefont {J.~I.}\ \bibnamefont
  {Cirac}},\ and\ \bibinfo {author} {\bibfnamefont {M.~B.}\ \bibnamefont
  {Hastings}},\ }\bibfield  {title} {\bibinfo {title} {{Strong and Weak
  Thermalization of Infinite Nonintegrable Quantum Systems}},\ }\href
  {https://doi.org/10.1103/PhysRevLett.106.050405} {\bibfield  {journal}
  {\bibinfo  {journal} {Phys. Rev. Lett.}\ }\textbf {\bibinfo {volume} {106}},\
  \bibinfo {pages} {050405} (\bibinfo {year} {2011})}\BibitemShut {NoStop}%
\bibitem [{\citenamefont {Zueco}\ \emph {et~al.}(2009)\citenamefont {Zueco},
  \citenamefont {Galve}, \citenamefont {Kohler},\ and\ \citenamefont
  {H\"anggi}}]{ZDG09}%
  \BibitemOpen
  \bibfield  {author} {\bibinfo {author} {\bibfnamefont {D.}~\bibnamefont
  {Zueco}}, \bibinfo {author} {\bibfnamefont {F.}~\bibnamefont {Galve}},
  \bibinfo {author} {\bibfnamefont {S.}~\bibnamefont {Kohler}},\ and\ \bibinfo
  {author} {\bibfnamefont {P.}~\bibnamefont {H\"anggi}},\ }\bibfield  {title}
  {\bibinfo {title} {{Quantum router based on ac control of qubit chains}},\
  }\href {https://doi.org/10.1103/PhysRevA.80.042303} {\bibfield  {journal}
  {\bibinfo  {journal} {Phys. Rev. A}\ }\textbf {\bibinfo {volume} {80}},\
  \bibinfo {pages} {042303} (\bibinfo {year} {2009})}\BibitemShut {NoStop}%
\bibitem [{\citenamefont {De~Chiara}\ \emph {et~al.}(2012)\citenamefont
  {De~Chiara}, \citenamefont {Lepori}, \citenamefont {Lewenstein},\ and\
  \citenamefont {Sanpera}}]{DGL12}%
  \BibitemOpen
  \bibfield  {author} {\bibinfo {author} {\bibfnamefont {G.}~\bibnamefont
  {De~Chiara}}, \bibinfo {author} {\bibfnamefont {L.}~\bibnamefont {Lepori}},
  \bibinfo {author} {\bibfnamefont {M.}~\bibnamefont {Lewenstein}},\ and\
  \bibinfo {author} {\bibfnamefont {A.}~\bibnamefont {Sanpera}},\ }\bibfield
  {title} {\bibinfo {title} {{Entanglement Spectrum, Critical Exponents, and
  Order Parameters in Quantum Spin Chains}},\ }\href
  {https://doi.org/10.1103/PhysRevLett.109.237208} {\bibfield  {journal}
  {\bibinfo  {journal} {Phys. Rev. Lett.}\ }\textbf {\bibinfo {volume} {109}},\
  \bibinfo {pages} {237208} (\bibinfo {year} {2012})}\BibitemShut {NoStop}%
\bibitem [{\citenamefont {Simon}\ \emph {et~al.}(2011)\citenamefont {Simon},
  \citenamefont {Bakr}, \citenamefont {Ma}, \citenamefont {Preiss},\ and\
  \citenamefont {Greiner}}]{SC1}%
  \BibitemOpen
  \bibfield  {author} {\bibinfo {author} {\bibfnamefont {J.}~\bibnamefont
  {Simon}}, \bibinfo {author} {\bibfnamefont {W.~S.}\ \bibnamefont {Bakr}},
  \bibinfo {author} {\bibfnamefont {M.~E. T.~R.}\ \bibnamefont {Ma}}, \bibinfo
  {author} {\bibfnamefont {P.~M.}\ \bibnamefont {Preiss}},\ and\ \bibinfo
  {author} {\bibfnamefont {M.}~\bibnamefont {Greiner}},\ }\bibfield  {title}
  {\bibinfo {title} {Quantum simulation of antiferromagnetic spin chains in an
  optical lattice},\ }\href {https://doi.org/10.1038/nature09994} {\bibfield
  {journal} {\bibinfo  {journal} {Nature}\ }\textbf {\bibinfo {volume} {472}},\
  \bibinfo {pages} {307} (\bibinfo {year} {2011})}\BibitemShut {NoStop}%
\bibitem [{\citenamefont {Greif}\ \emph {et~al.}(2013)\citenamefont {Greif},
  \citenamefont {Uehlinger}, \citenamefont {Jotzu}, \citenamefont {Tarruell},\
  and\ \citenamefont {Esslinger}}]{SC2}%
  \BibitemOpen
  \bibfield  {author} {\bibinfo {author} {\bibfnamefont {D.}~\bibnamefont
  {Greif}}, \bibinfo {author} {\bibfnamefont {T.}~\bibnamefont {Uehlinger}},
  \bibinfo {author} {\bibfnamefont {G.}~\bibnamefont {Jotzu}}, \bibinfo
  {author} {\bibfnamefont {L.}~\bibnamefont {Tarruell}},\ and\ \bibinfo
  {author} {\bibfnamefont {T.}~\bibnamefont {Esslinger}},\ }\bibfield  {title}
  {\bibinfo {title} {Short-range quantum magnetism of ultracold fermions in an
  optical lattice},\ }\href {https://doi.org/10.1126/science.1236362}
  {\bibfield  {journal} {\bibinfo  {journal} {Science}\ }\textbf {\bibinfo
  {volume} {340}},\ \bibinfo {pages} {1307} (\bibinfo {year}
  {2013})}\BibitemShut {NoStop}%
\bibitem [{\citenamefont {Pham}\ \emph {et~al.}(2011)\citenamefont {Pham},
  \citenamefont {Malinowski},\ and\ \citenamefont {Bartczak}}]{PNM11}%
  \BibitemOpen
  \bibfield  {author} {\bibinfo {author} {\bibfnamefont {N.}~\bibnamefont
  {Pham}}, \bibinfo {author} {\bibfnamefont {A.}~\bibnamefont {Malinowski}},\
  and\ \bibinfo {author} {\bibfnamefont {T.}~\bibnamefont {Bartczak}},\
  }\bibfield  {title} {\bibinfo {title} {Comparative study of derivative free
  optimization algorithms},\ }\href {https://doi.org/10.1109/tii.2011.2166799}
  {\bibfield  {journal} {\bibinfo  {journal} {IEEE Trans. Industr. Inform.}\
  }\textbf {\bibinfo {volume} {7}},\ \bibinfo {pages} {592} (\bibinfo {year}
  {2011})}\BibitemShut {NoStop}%
\bibitem [{\citenamefont {Hogben}\ \emph {et~al.}(2010)\citenamefont {Hogben},
  \citenamefont {Hore},\ and\ \citenamefont {Kuprov}}]{HHH10}%
  \BibitemOpen
  \bibfield  {author} {\bibinfo {author} {\bibfnamefont {H.}~\bibnamefont
  {Hogben}}, \bibinfo {author} {\bibfnamefont {P.}~\bibnamefont {Hore}},\ and\
  \bibinfo {author} {\bibfnamefont {I.}~\bibnamefont {Kuprov}},\ }\bibfield
  {title} {\bibinfo {title} {Strategies for state space restriction in densely
  coupled spin systems with applications to spin chemistry},\ }\href
  {https://doi.org/10.1063/1.3398146} {\bibfield  {journal} {\bibinfo
  {journal} {J. Chem. Phys.}\ }\textbf {\bibinfo {volume} {132}},\ \bibinfo
  {pages} {174101} (\bibinfo {year} {2010})}\BibitemShut {NoStop}%
\bibitem [{\citenamefont {Ladd}\ \emph {et~al.}(2010)\citenamefont {Ladd},
  \citenamefont {Jelezko}, \citenamefont {Laflamme}, \citenamefont {Nakamura},
  \citenamefont {Monroe},\ and\ \citenamefont {O'Brien}}]{LTD10}%
  \BibitemOpen
  \bibfield  {author} {\bibinfo {author} {\bibfnamefont {T.~D.}\ \bibnamefont
  {Ladd}}, \bibinfo {author} {\bibfnamefont {F.}~\bibnamefont {Jelezko}},
  \bibinfo {author} {\bibfnamefont {R.}~\bibnamefont {Laflamme}}, \bibinfo
  {author} {\bibfnamefont {Y.}~\bibnamefont {Nakamura}}, \bibinfo {author}
  {\bibfnamefont {C.}~\bibnamefont {Monroe}},\ and\ \bibinfo {author}
  {\bibfnamefont {J.~L.}\ \bibnamefont {O'Brien}},\ }\bibfield  {title}
  {\bibinfo {title} {Quantum computers},\ }\href
  {https://doi.org/10.1038/nature08812} {\bibfield  {journal} {\bibinfo
  {journal} {Nature}\ }\textbf {\bibinfo {volume} {464}},\ \bibinfo {pages}
  {45} (\bibinfo {year} {2010})}\BibitemShut {NoStop}%
\end{thebibliography}
\end{document}